\documentclass[12pt]{iopart}
\usepackage{iopams,subfigure,graphicx}
\usepackage{citesort}
\expandafter\let\csname equation*\endcsname\relax
\expandafter\let\csname endequation*\endcsname\relax
\usepackage{amsmath}
\usepackage{color}

\begin{document}
\title{Dark-soliton-like excitations in the Yang-Gaudin gas of attractively interacting fermions}
\author{Sophie S.~Shamailov, Joachim Brand}
\address{Dodd-Walls Centre for Photonics and Quantum Technology, Centre for Theoretical Chemistry and Physics, New Zealand Institute for Advanced Study,
Massey University, Private Bag 102904 NSMC, Auckland 0745, New Zealand}
\ead{J.Brand@massey.ac.nz}
\vspace{10pt}
\date{\today}
\begin{abstract}
Yrast states are the lowest energy states at given non-zero momentum and provide a natural extension of the concept of dark solitons to strongly-interacting one-dimensional quantum gases. Here we study the yrast states of the balanced spin-$\frac{1}{2}$ Fermi gas with attractive delta-function interactions in one dimension with the exactly solvable Yang-Gaudin model. The corresponding Bethe-ansatz equations are solved for finite particle number and in the thermodynamic limit. Properties corresponding to the soliton-like nature of the yrast excitations are calculated including the missing particle number, phase step, and inertial and physical masses. The inertial to physical mass ratio, which is related to the frequency of oscillations in a trapped gas, is found to be unity in the limits of strong and weak attraction and falls to $\approx 0.78$ in the crossover regime. This result is contrasted by one-dimensional mean field theory, which predicts a divergent mass ratio in the weakly attractive limit. By means of an exact mapping our results also predict the existence and properties of dark-soliton-like excitations in the super Tonks-Girardeau gas. The prospects for experimental observations are briefly discussed.
\end{abstract}
\vspace{2pc}
\submitto{\NJP}
%\maketitle
%\tableofcontents
%\newpage
\section{\label{intro}Introduction}
Solitons are ubiquitous nonlinear wave phenomena that appear in many different fields of physics \cite{Dauxois2006}. They often appear in the context of fluid dynamics or the nonlinear mean-field theories of quantum systems \cite{Kevrekidis2008}. A very interesting open question is to what degree solitary wave phenomena can survive in low-dimensional and strongly correlated quantum systems, where mean-field theories or fluid-dynamic-like descriptions are inappropriate due to the dominance of quantum effects.

Being characterized by a localized and non-dispersing depression in the particle number density and a step variation in the superfluid phase, dark solitons \cite{Tsuzuki1971} have been observed in Bose-Einstein condensates (BECs) of repulsively-interacting ultra-cold atoms \cite{Burger1999,Denschlag2000,Becker2008}. Ultra-cold gases of fermionic atoms \cite{Kinast2005,Zwierlein2005,Liao2010,Navon2010,Joseph2011} have extended the physics of gaseous superfluids beyond weakly-interacting BECs  by realizing the crossover from a BEC of tightly-bound molecules to a fermionic superfluid with long-range pair correlations described by BCS theory \cite{Giorgini2008,Zwerger2012}. Dark solitons have been predicted \cite{Dziarmaga2004b,Antezza2007,Liao11pr:FermiSolitons,Scott2011,Spuntarelli2011,Scott2012} and observed  \cite{Ku} in the crossover regime. In one-dimensional (1D) Fermi superfluids with strong spin-orbit coupling, dark solitons have been predicted to carry Majorana fermions as quasiparticles \cite{Xu2014,Liu2015,Zou2015}. 

An interesting feature of dark solitons is that they exhibit oscillatory motion around the position of maximum density in a trapped gas. The frequency of oscillation $\Omega$  is universally related to the ratio of the inertial mass $m_I$ and the physical mass $m_P$ under the condition of small oscillations \cite{Konotop2004,Scott2011,Liao11pr:FermiSolitons}
\begin{equation} \label{eq:massratio}
\frac{m_I}{m_P} = \left(\frac{\omega_\mathrm{trap}}{\Omega}\right)^2 ,
\end{equation}
where  $\omega_\mathrm{trap}$ is the frequency of the harmonic trapping potential. The mass ratio $m_I/m_P$ is a non-trivial characteristic of the underlying many-body physics of the medium. For (tightly confined) atomic BECs the ratio $m_I/m_P =2$ was predicted from the Gross-Pitaevskii equation in the nonlinear Thomas-Fermi limit \cite{Busch2000,Konotop2004}, and measured experimentally in \cite{Becker2008,Weller2008}. In the BEC to BCS crossover in a Fermi superfluid, the ratio was predicted to increase from 2 in the BEC regime to 3 in the unitary regime and grow even larger towards the BCS regime in \cite{Liao11pr:FermiSolitons,Scott2011}. Unfortunately, the dark solitons observed in experiments  decayed before their oscillation frequency could be measured by means of the snaking instability due to the three-dimensional (3D) nature of the trapped gas  \cite{Ku,Cetoli2013,Mateo2014}. Instead, a solitonic vortex \cite{brand01a,Brand2002,Komineas2003} was observed  \cite{Yefsah,Ku2014}, that owes its large mass ratio  $m_I/m_P$ (in the hundreds) to hydrodynamic effects \cite{Ku2014,Mateo2015a}.

A natural way to stabilize the dark soliton  is to reduce the dimensionality by confining the gas to a tight wave-guide-like trap \cite{Muryshev1999}. In order to suppress the snaking instability for the crossover Fermi superfluid, a transverse confinement length scale comparable to the mean particle spacing is required \cite{Cetoli2013,Mateo2014}, which means entering a 1D regime. An analytically solvable 1D mean-field model for moving dark solitons in the BCS regime  predicts that the mass ratio $m_I/m_P$ diverges to $+\infty$ in the BCS limit of weak attractive interactions \cite{Efimkin2014} for a quasi one-dimensional channel (whereas  $m_I/m_P \to -\infty$ for a strict 1D version of their model). These results are consistent with the predictions from 3D mean-field theory \cite{Liao11pr:FermiSolitons,Scott2011}, where  $m_I/m_P$ increases towards the BCS regime. As we will show in this work, however, exact solutions of the Yang-Gaudin model predict that   $m_I/m_P$ approaches unity in the weakly-coupled BCS limit for a 1D Fermi gas. The validity of mean-field theory is generally questionable in strongly-correlated regimes where no small parameter exists \footnote{Note that the initial discrepancy between the experiment \cite{Yefsah} and theory \cite{Liao11pr:FermiSolitons,Scott2011} has since been resolved \cite{Ku2014,Ku}.}, and in the strongly-confined 1D regime where phase fluctuations prevent true long-range superfluid order according to the Mermin-Wagner Hohenberg theorem \cite{Mermin1966,hohenberg67}. Nevertheless, the use of BCS-like mean-field theory in the weakly-attractive 1D Fermi gas had previously been supported by the comparison of bulk properties with exact solutions of the Yang-Gaudin model \cite{Liu2007b}.

\begin{figure}[htbp]
\begin{center}
\includegraphics[width=0.7\columnwidth]{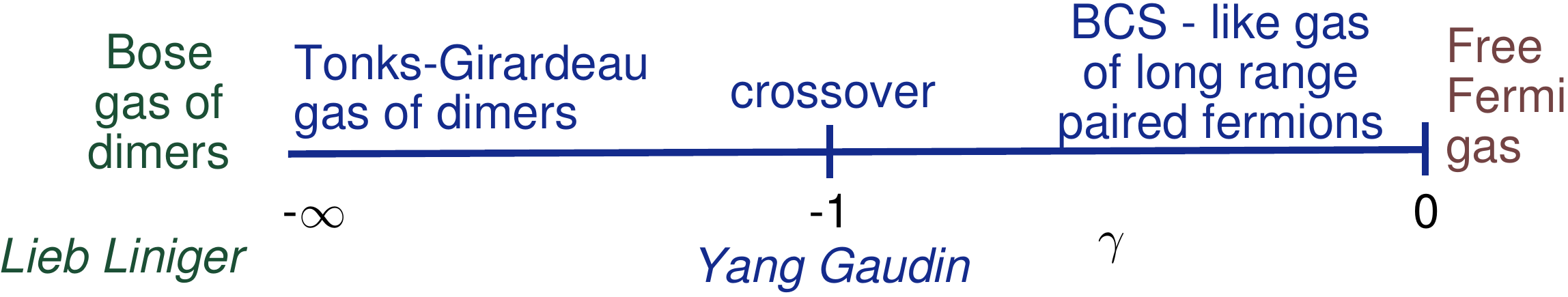}
\caption{\label{YGschem} Schematic of the spin--$\frac{1}{2}$ Fermi gas with attractive interactions in a one-dimensional wave guide. The physical regimes of the many-body physics are determined by the dimensionless coupling parameter $\gamma = 2/a_\mathrm{1D} n_0 = gm/n_0\hbar^2 = c/n_0$.}
\end{center}
\end{figure}

Here we approach the tightly confined 1D Fermi gas from the exactly solvable Yang-Gaudin model \cite{Yang,Gaudin} of spin-$\frac{1}{2}$ fermions with attractive $\delta$-function interactions. Using a purely 1D model with contact interactions is well justified for an ultra-cold Fermi gas in the low density regime, where the relevant energy scale, e.g.\ the Fermi energy of the 1D gas, is small compared to the spacing of transverse excitation levels of the wave-guide trap \cite{Fuchs2004,Tokatly2004,Astrakharchik2004b,Mora2005}. While the two-particle problem in a wave-guide trap always has a bound state \cite{Bergeman2003}, the many-body problem of spin-$\frac{1}{2}$ fermions knows two distinct regimes that are separated by a confinement-induced resonance, where the 1D scattering length $a_\mathrm{1D}$ passes through zero. The regime of $a_\mathrm{1D} < 0$ is well described by the Yang-Gaudin model of spin-$\frac{1}{2}$ fermions with attractive contact interactions while positive scattering lengths $a_\mathrm{1D} > 0$ lead to a gas of repulsively interacting bosonic dimers described by the Lieb-Liniger model \cite{LL1,LL2}. The situation is sketched in figure \ref{YGschem} with reference to the relevant dimensionless coupling parameter $\gamma = 2/a_\mathrm{1D} n_0$ of the Yang-Gaudin model. Here, $n_0=N/L$ is the particle number density, $N$ the total number of fermions and $L$ the length of the system. 

The Yang-Gaudin model with attractive interactions ($\gamma<0$) has two-particle bound states, dimers, with length scale $\sim |a_\mathrm{1D}|$. The coupling constant $\gamma = 2/a_\mathrm{1D} n_0$ thus provides the ratio of the  mean particle spacing $n_0^{-1}$ to the dimer size and $|\gamma|\ll 1$ is a regime of Cooper-pair-like large dimers. An interesting crossover regime appears at  $|\gamma| \approx 1$, where the length scales are comparable, while the dimers are tightly bound for $\gamma \to -\infty$. This is the position of the confinement-induced resonance, where the dimers are tightly bound but also the dimer-dimer interactions diverge \cite{Mora2005}. The many-body physics becomes that of a gas of impenetrable bosons, known as a Tonks-Girardeau gas \cite{olshanii98,girardeau60}. In this regime, the energy spectrum becomes that of $N/2$ spinless fermions with mass $2m$, where $m$ is the mass of the original fermionic atoms. In fact, there is a one-to-one mapping between the impenetrable Bose gas (here referring to dimers) and non-interacting spinless fermions (of mass $2m$) \cite{girardeau60}. In this sense both limits of the attractive Yang Gaudin model are non-interacting fermions and the interesting correlated physics is expected around $|\gamma| \approx 1$. 

There is a close relation between the attractively interacting Fermi gas with a balanced population of spin-up and -down fermions and the repulsively interacting Bose gas. The 1D Bose gas with repulsive $\delta$-function interactions is known as the Lieb-Liniger model and its ground state energy and excitation spectrum were solved for in \cite{LL1,LL2}. An important finding was a gapless branch of excitations with lower energy than the gapless branch of phonon excitations (similar to the Bogoliubov phonons of the 3D Bose gas). Lieb called these \emph{type-II} elementary excitations \cite{LL2}. They are part of the set of \emph{yrast} states, which refers to the lowest energy state at given momentum \footnote{The term \emph{yrast} (Swedish for ``dizziest'') was originally used in nuclear physics to refer to high-spin but low energy nuclear states. In the present context, where momentum in a periodic box is equivalent to angular momentum on a ring, the term was first used in \cite{Kanamoto2008}.}.

By now there is mounting evidence for the association of the yrast excitations of the Lieb-Liniger model to dark solitons known to exist in the 3D BEC and found mathematically in the nonlinear Schr\"odinger equation \cite{Kulish1976,Ishikawa1980,Kanamoto2008,Kanamoto2009,Kanamoto2010,Fialko2012b,Syrwid2015,Kaminishi2011,Sato2012a,Sato2016}. References  \cite{Kulish1976,Ishikawa1980} first identified the energy--momentum dispersion relation of dark solitons with Lieb's type-II branch in the weak coupling limit, which was later extended to multi-soliton solutions \cite{Kanamoto2008,Kanamoto2009,Kanamoto2010}. Reference \cite{Fialko2012b} further strengthened the connection by establishing how the localized dark solitons of mean-field theories (and experimental observations) emerge from a superposition of the translationally invariant yrast states. Syrwid and Sacha went on to demonstrate that localized density depressions typical of dark solitons emerge at random positions from the translationally invariant 
yrast states under a single-shot measurement procedure, where the position of the atoms is detected one-by-one \cite{Syrwid2015}.

The association between dark solitons and yrast excitations in the Lieb-Liniger model opens the door to a deeper investigation of solitonic properties of yrast states in the more strongly-interacting regimes of 1D quantum gases, outside the validity of mean-field theory. Astrakharchik and Pitaevskii have studied the (type-II) dispersion relation of yrast excitations in the Lieb-Liniger model throughout the crossover from a weakly-interacting Bose gas to the strongly-correlated Tonks-Girardeau regime and deduced a number of quantities of interest \cite{Astrakharchik2012}. These include the inertial and physical masses. One important finding is that the effective particle number $m_P/m$ diverges, and thus becomes macroscopic, in the limit of small coupling constant $\gamma$, while corresponding to a single boson hole in the Tonks-Girardeau regime of large $\gamma$. The effective particle number  $m_P/m < 0$ (called $N_s$ in \cite{Astrakharchik2012}) is  closely related to the number of atoms $N_d < 0$ in the dark soliton, i.e.\ the number of atoms that have to be removed from the background density to create an observed density depression. Another result of \cite{Astrakharchik2012} is that the inertial-to-physical mass ratio $m_I/m_P$, which relates to the oscillation frequency of the soliton in a harmonically trapped gas, smoothly decreases from the value of 2 in the weak coupling limit of the 1D Bose gas (where it agrees with the mean-field prediction), to 1 in the Tonks-Girardeau regime.

In this work we extend the study of yrast excitations to the spin-$\frac{1}{2}$ Fermi gas with attractive interactions within the Yang-Gaudin model. We identify yrast excitations in the finite system as dimer-hole excitations in the strongly-attractive regime and clarify their continuous connection to single fermion holes in the non-interacting limit by solving the Bethe-ansatz equations directly and with the aid of a string hypothesis for fermion pairs in the strongly attractive regime. We then  solve the integral equations for dimer holes, which govern the yrast dispersion relation in the thermodynamic limit. From these, a number of physical quantities including the missing particle number, the phase step, and the inertial and physical masses are calculated. We find that the missing particle number varies from one to two missing fermions as we tune from the Tonks-Girardeau gas of dimers to the BCS limit, while both the physical and the inertial masses correspondingly change from $-2m$ to $-m$. Interestingly, the mass ratio $m_I/m_P$ is 1 in both limits and reaches a minimum value of about $0.78$ in the crossover regime $\gamma \approx -1.3$ (see figure \ref{maF2_single}). This contrasts sharply with the results of 1D mean field theory \cite{Efimkin2014}, where the mass ratio diverges in the BCS limit.

There is a close relation between the attractive Fermi gas and the super Tonks-Girardeau gas, a highly excited gaseous phase of attractive bosons in 1D  \cite{Astrakharchik2004,Astrakharchik2005,Haller2009,Batchelor2005,Batchelor2010}. Even though  attractive bosons form cluster-like bound states or bright solitons in the ground state, a metastable gaseous phase can be supported by kinetic energy pressure in a similar way that Fermi pressure stabilizes the attractive Fermi gas. Due to a one-to-one correspondence of the Bethe-ansatz solutions of the Yang-Gaudin model to the Lieb-Liniger model established by Chen \emph{et al.} \cite{Batchelor2010}, our results carry over to the super Tonks-Girardeau gas where they predict the existence of dark soliton-like features, which is quite remarkable. Due to the stability properties of yrast states we may expect the life time of these states to be comparable to that of the ``ground-state'' super Tonks-Girardeau gas.

This paper is organized as follows: \Sref{YGmodel} introduces the Yang-Gaudin model and the corresponding Bethe ansatz equations for finite particle number. The structure of the ground and yrast exctited states, as well as the methods for solving the equations, are discussed in both the strongly and weakly interacting regimes, and the excitation spectra are computed. \Sref{TDL_eqns} then deals with the thermodynamic limit dispersion relations. After discussing the ground state thermodynamic-limit equations in \sref{TDL_aF2gs}, the equations for the yrast dispersion relations are given in \sref{TDL_aF2s}, the missing particle number and phase step are calculated in \sref{Ns} and the inertial and physical masses in\sref{Masses}. Conclusions are drawn in \sref{disc_conc} and three appendices provide technical details regarding the numerical solution of the finite-system Bethe ansatz equations (\ref{TechDetails}), an outline of the derivation of the excited-state thermodynamic limit integral equations (\ref{TDLhowto}), and details pertaining to the numerical computation of the missing particle number (\ref{Ns_aF2s}), respectively.
\section{\label{YGmodel}Yang-Gaudin model for finite fermion number}
The Yang-Gaudin model describes a gas of spin-1/2 fermions, confined to a 1D box with periodic boundary conditions and interacting via a two-body $\delta$-function potential. The Hamiltonian takes the form
\begin{equation}
\label{Ham}
H = -\frac{\hbar^2}{2m}\sum\limits_{j=1}^N \frac{d^2}{dx_j^2} + g\sum\limits_{\left\langle i,j\right\rangle} \delta\left(x_i-x_j\right),
\end{equation}
where in the second term the sum is over all pairs counted once.  There are $N$ fermions in total, $M\leq N/2$ of which are spin-up and the rest are spin-down. Furthermore, $m$ is the mass of each particle, and $L$ is the length of the box. The model is relevant to ultra-cold fermionic atoms in two hyperfine states confined to a 1D wave guide in the low density limit \cite{Fuchs2004,Tokatly2004,Astrakharchik2004b,Mora2005}. The 1D interaction constant $g$ can be related to the 3D scattering length $a$ of the atoms, the wave guide trap frequency $\omega_\perp$ and length scale $a_\perp = \sqrt{\hbar/m\omega_\perp}$  by $g = 2\hbar\omega_\perp a (1-A a/a_\perp)^{-1}$ with $A\approx 1.0326$ \cite{olshanii98}. The coupling constant diverges and changes sign at a confinement-induced resonance where $a/a_\perp = A^{-1}$. We are here considering the case of attractive interaction where $g<0$. It is convenient to write the interaction constant as $g=\frac{\hbar^2}{m}c$ where $c={2}/{a_{1D}}$ and $a_{1D}$ is the 1D scattering length \cite{Batchelor2005II}. Let us also introduce the 1D density, $n_0=N/L$, and $\gamma=c/n_0$ which is a useful dimensionless parameter both in the finite system and in the thermodynamic limit.
\subsection{\label{YGinteracting}Bethe ansatz equations in exponential form}
\begin{figure}[htbp]
\begin{center}
\includegraphics[width=6in]{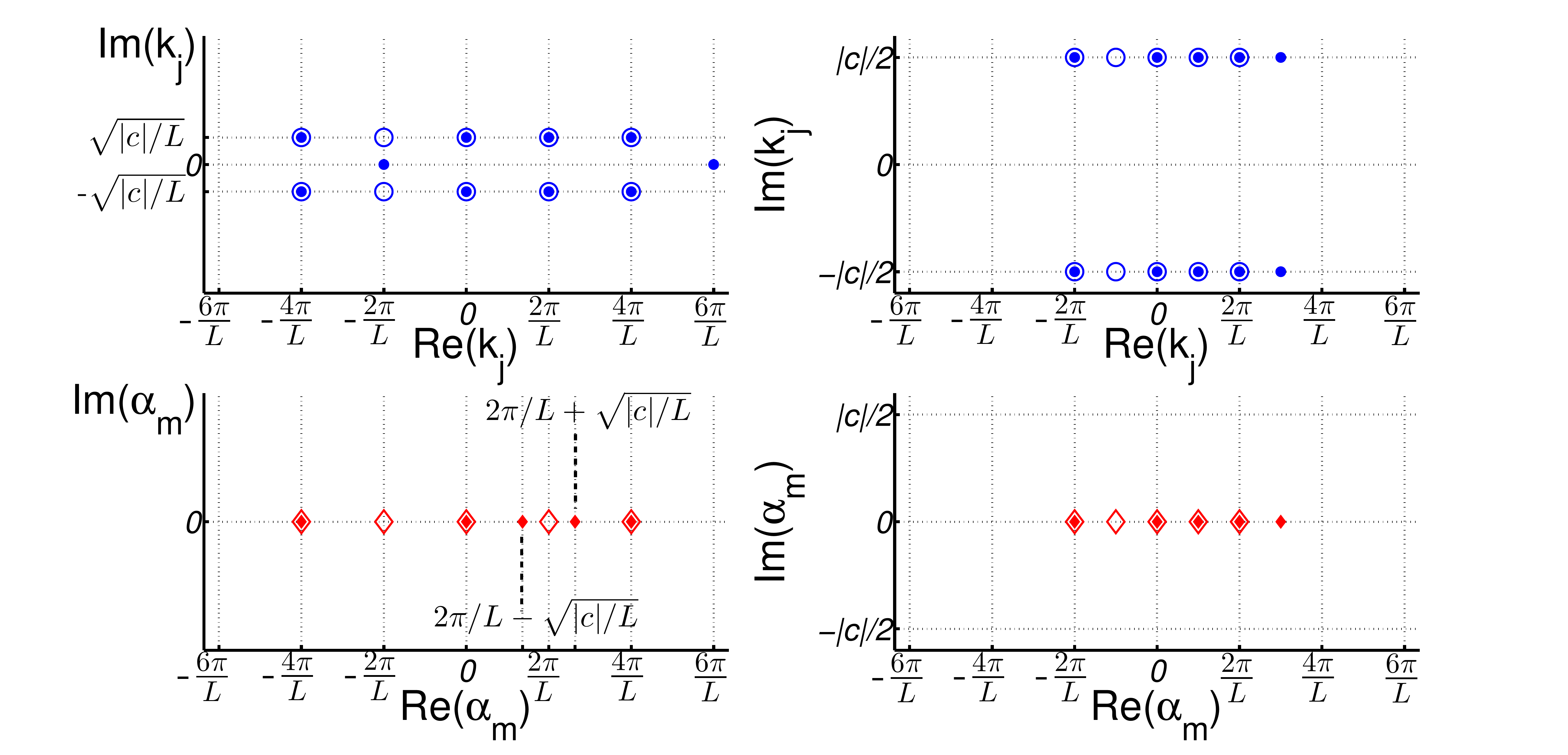}
\caption{\label{diag_1f} A schematic representation of the rapidities on the complex plane for particular yrast states with $N=10$ spin-$\frac{1}{2}$ fermions and a balanced spin population ($M=5$). Charge rapidities $\{k_j\}$ are shown on top and spin rapidities $\{\alpha_m\}$ on the bottom, while the left panels refer to weak ($c\lesssim 0$) and the right panels to strong attractive interactions ($c\rightarrow-\infty$). Empty symbols show the ground state ($P_\mathrm{tot}=0$) and resemble filled Fermi spheres. Filled-in symbols show an yrast excited state with $P_\mathrm{tot}= 8\pi\hbar/L$. In the strongly attractive case (right), this excitation is obtained by creating a hole in the ``Fermi-sphere'' of dimers, i.e.\ moving the set of two charge and one spin rapidities from quasi-momentum $-\pi/L$  to the outside of the sphere, at $3\pi/L$. In the weekly-interacting regime (left), the corresponding yrast excitation is obtained by taking a single fermion at $-2\pi/L$ out of the Fermi sphere. Note that this state is a non-degenerate singlet excitation while the related triplet excitations have higher energy for nonzero $c$.}
\end{center}
\end{figure}

The Bethe-ansatz solution of the Yang-Gaudin model \eref{Ham} consists of superpositions of plane waves for the many-body wave function \cite{Sutherland2004}. These are subject to boundary conditions where particles interact and at the box boundaries, as well as fermionic symmetry constraints. The solutions are uniquely determined by a set of variables with the dimension of wave numbers known as rapidities. They have to satisfy the Bethe ansatz equations in exponential form \cite{Yang,Gaudin}:
\begin{eqnarray}
\label{expF2a}
\exp(ik_j L) &=& \prod\limits_{n=1}^{M}\frac{k_j-\alpha_n+ic/2}{k_j-\alpha_n-ic/2},\\
\label{expF2b}
\prod\limits_{j=1}^{N}\frac{\alpha_m-k_j+ic/2}{\alpha_m-k_j-ic/2} &=& -\prod\limits_{n=1}^{M}\frac{\alpha_m-\alpha_n+ic}{\alpha_m-\alpha_n-ic}.
\end{eqnarray}
The charge rapidities $k_j$ can be thought of as the quasi-momenta of the fermions. They completely determine the total momentum and energy of the system
\begin{eqnarray}
P_\mathrm{tot} &=& \hbar\sum\limits_{j=1}^{N} k_j,\\ \label{eq:energy}
E_\mathrm{tot} &=& \frac{\hbar^2}{2m}\sum\limits_{j=1}^{N} k_j^2.
\end{eqnarray}
The spin rapidities $\alpha_m$ are auxiliary variables and are present due to the spin degree of freedom. The $\alpha_m$'s do not contribute to the energy or momentum but must be solved for as they are coupled to the $k_j$'s. There are infinitely many different sets of rapidities that solve \eref{expF2a} and  \eref{expF2b} and each one corresponds to an eigenstate of the Hamiltonian \eref{Ham}. In this work we are interested in the yrast states, i.e.\ the states with the lowest energy $E_\mathrm{tot}$ at given momentum $P_\mathrm{tot}$. We also restrict ourselves to balanced populations of spin-up and -down particles, i.e.\ $N=2M$. The rapidities for yrast states can be easily identified in the weak and strong interaction limits, where simple analytic solutions to  \eref{expF2a} and  \eref{expF2b} are known. The yrast solutions for finite interaction strength can then be found by continuity. Examples of rapidities for yrast states are shown in \fref{diag_1f}.

A particular feature due to the periodic boundary conditions is that a new set of rapidities solving \eref{expF2a} and  \eref{expF2b} can be generated from an existing one by adding $2\pi/L$ (or an integer mutiple) to every $k_j$ and $\alpha_m$, as is easily seen from the equations. This changes the momentum to $P_\mathrm{tot}' = P_\mathrm{tot} +2\pi\hbar n_0$, where $n_0 = N/L$ and the energy to $E_\mathrm{tot}'=E_\mathrm{tot} + 2\pi\hbar P_\mathrm{tot}/mL + (2\pi\hbar n_0)^2/2mN$. Physically this corresponds to a Galilean boost of the whole system by the \emph{umklapp} momentum $2\pi\hbar n_0 \equiv 4 p_F$, while the internal structure of the many-body state is unchanged \cite{LL1}. Here, $p_F = \pi \hbar n_0/2$ is the Fermi momentum of the non-interacting spin-$\frac{1}{2}$ Fermi gas. It is thus sufficient to consider yrast states in the interval $0 \le P < 4 p_F$. 

The Bethe ansatz equations \eref{expF2a} and  \eref{expF2b} can be  solved directly by numerical methods. However it is only practical to do so for small particle number and in a regime of very weak interactions. The situation is similar to that in the Lieb-Liniger model with attractive interactions, where bound states are formed as well \cite{Sykes2007}. Technical details regarding the numerical solutions are given in \ref{TechDetails}.
\subsection{\label{GSexact}Weak interactions}
In the case of weak interactions $|\gamma|\ll 1$, an analytical result is available for the rapidities of the ground state 
\cite{GaudinThesis}. Define the set of integers $q_m$ as
\begin{equation}\label{eq:weak}
q_m = -\frac{M+1}{2}+m\ \left(+\frac{1}{2}\right),\ \ \ m = 1,2,\ldots,M,\\
\end{equation}
where the term in brackets is only included for even $M$. Then $\alpha_m=\frac{2\pi}{L}q_m$ and $k_{2m,2m-1} = \frac{2\pi}{L}q_m\pm\sqrt{\frac{c}{L}}$. An illustration of the rapidities for the case of a non-degenerate ground state with odd $M$ is shown in \fref{diag_1f} (left, empty symbols). For $c<0$ pairs of $k_j$'s are complex conjugates and there is a limited range where the exponential equations can be solved in practice.

The energy for weak interactions can be obtained by explicit evaluation of the sum \eref{eq:energy} \cite{Guan2013}. In terms of the dimensionless coupling parameter $\gamma = c/n_0$ we obtain (assuming odd $M$)
\begin{equation}
\label{Eapproxw}
 E^\mathrm{weak}_\mathrm{GS} = \frac{\hbar^2 n_0^2}{2m} \left(M\gamma + \frac{\pi^2}{12}\frac{N^2-4}{N}\right) .
\end{equation}
This approximation is compared to the exact results in the bottom panel of Fig.~\ref{F2_grnd}. For $c=0$ this becomes simply the expression for the energy of a gas of non-interacting spin-$\frac{1}{2}$ fermions.

Yrast excitations in the small $|c|$ limit are obtained by creating a single-fermion hole, which has further implications. When a single $k_j$ is moved from $\frac{2\pi}{L}q_m\pm\sqrt{c/L}$ out of the Fermi spheres to $\frac{2\pi}{L}(q_M+1)$, its partner moves from $\frac{2\pi}{L}q_m\mp\sqrt{c/L}$ to $\frac{2\pi}{L}q_m$ and $\alpha_m$ shifts to half-way between its associated $k_j$'s. For half of the excitations, this means $\alpha_m$ becomes a half-integer multiple of $2\pi/L$ and for the other half, $\alpha_m$ is expelled to a position already occupied by another $\alpha$. In the latter case, these two $\alpha$'s split away from each other by $\pm \sqrt{|c|/L}$. An illustrative example of this type of excitation is given in the left column of \fref{diag_1f}, 
where we use $N=10, M=5$ and $q_m=-1$. To summarize, the real parts of the $k_j$'s are determined by the free fermion limit while the splitting in the imaginary part is the same as for the ground state; the $\alpha_m$'s remain real-valued and sit at the average of their corresponding $k_j$ pairs except where a degeneracy would result, in which case the degeneracy is lifted by splitting along the real axis.
\subsection{\label{GS_SHc<0}String hypothesis}
For stronger attractive interaction, as the value of $c$ becomes increasingly negative,  the real part of pairs of $k_j$ values becomes equal to the value of $\alpha_m$ while the imaginary parts become exponentially close to $\pm i\frac{c}{2}$. This causes divergent terms  in \eref{expF2a} and  \eref{expF2b} through vanishing denominators and provides a complication for numerical root finding. In this case, approximate equations can be derived using the so-called  \textit{string-hypothesis}, where the imaginary parts are replaced by the limiting values \cite{GaudinThesis}. Specifically for the ground state, $2M$ of the $k_j$'s  are taken to be $\alpha_m \pm i\frac{c}{2}$, and the remaining $N-2M$ real $k_j$'s are left as unknown variables. The paired charge rapidities can be eliminated from the equations by substitution, with all remaining variables (the $\alpha_m$'s and the unpaired $k_j$'s) being real and distinct. It is then possible to take the logarithm of the Bethe-ansatz equations. Assuming that all fermions are paired, which is adequate for the balanced ground state and spin-conserving yrast excitations, this leads to \cite{Sutherland2004}:
\begin{eqnarray}
\label{logaF2b}
2\alpha_mL &=& 2\pi \ell_m + \sum\limits_{n=1}^M \theta(\alpha_m-\alpha_n),
\end{eqnarray}
where
\begin{equation}
\label{thetafnct}
\theta(k) = -2\tan^{-1}\left(\frac{k}{c}\right)
\end{equation}
is the two-body phase-shift function of the $\delta$-function potential. The $\ell_m$'s are quantum numbers that specify the state. These are either integers or half-integers, depending on the parity of $N$ and $M$, and they must be distinct. Since we know the relation between the $\alpha_m$'s and the $k_j$'s associated with them, it is easy to write the momentum and energy as
\begin{eqnarray}
P_\mathrm{tot} &=& \hbar\sum\limits_{m=1}^{M} 2\alpha_m,\\ \label{eq:energy2}
E_\mathrm{tot} &=& \frac{\hbar^2}{2m}\sum\limits_{m=1}^{M} \left(2\alpha_m^2 - \frac{c^2}{2}\right).
\end{eqnarray}
Note the total binding energy of $-\frac{\hbar^2}{2m}M\frac{c^2}{2}$. The quantum numbers for the balanced ground state are
\begin{equation}
\label{aF2lms}
\ell_m = -\frac{M+1}{2} + m\ (+ 1),\ \ \ m = 1,2,\ldots,M,\\
\end{equation}
with the term in brackets only included for even $M$. In order to solve the logarithmic string hypothesis equations it is easier to start from the $c\rightarrow-\infty$ limit where it suffices to guess $\alpha_m=\frac{\pi \ell_m}{L}$ in order to converge to the solution.

The structure of the ground state rapidities in the strongly attractive regime ($c\rightarrow -\infty$) is illustrated in \fref{diag_1f} on the right by empty symbols. Indeed, it can be seen from the logarithmic equation \eref{logaF2b} that  the $\alpha_m$'s approach $\frac{\pi}{L}\ell_m$ where the $\ell_m$'s are given by (\ref{aF2lms}) (and the corresponding charge rapidities approach $\frac{\pi}{L}\ell_m\pm i\frac{c}{2}$). The structure of the ground state is now that of a Tonks-Girardeau gas of dimers: The distribution of the $\alpha_m$ rapidities is that of a Fermi sphere with $M$ spinless fermions, each corresponding to a bound pair of elementary fermions. 

\begin{figure}[htbp]
\begin{center}
\includegraphics[width=6in]{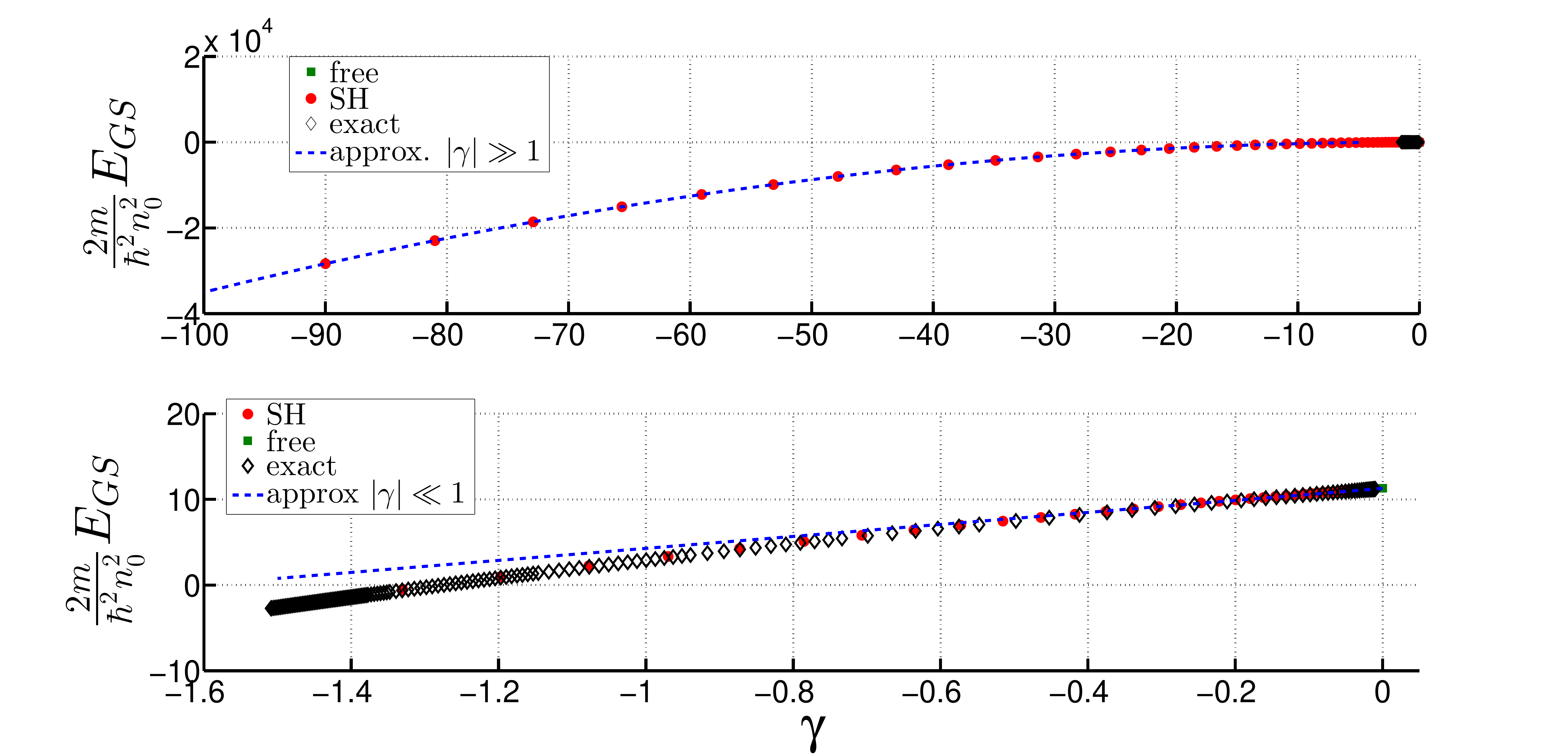}
\caption{\label{F2_grnd} Ground state energy as a function of coupling strength $\gamma$ for a system with $N=14, M=7$. Panel (b) shows a magnified version of the data from panel (a) around the region where exact results from the exponential equations are available in the attractive regime, and demonstrates that the string-hypothesis results compare well to the exact energy. Red circles show string-hypothesis results (labeled SH), black diamonds -- exact, green square -- free system. Shown as dashed blue lines are the approximate expressions (\ref{Eapproxs}) in (a) and (\ref{Eapproxw}) in (b).}
\end{center}
\end{figure}

The ground state energy can be obtained from \eref{eq:energy2} using the expressions for the rapidities in the case of strong interactions. We obtain the energy for a Tonks-Girardeau gas of $N/2$ bosonic dimers of mass $2m$ and binding energy $-\frac{\hbar^2 c^2}{4m}$, as
\begin{equation}
\label{Eapproxs}
 E_\mathrm{GS}^\mathrm{strong} =  \frac{\hbar^2 n_0^2}{2m} \left(-\gamma^2 \frac{N}{4} + \frac{\pi^2}{48}\frac{N^2-4}{N}\right).
\end{equation}
This expression is plotted in the top panel of figure \ref{F2_grnd} which shows the ground state energy as a function of coupling strength across the range $-100<\gamma<0$ for a system with $N=14, M=7$. In particular, the bottom panel allows us to compare the performance of the string hypothesis equations to the exact exponential equations in the region where $c<0$ and $|\gamma|\approx1$ or less. We see that the energy of the ground state is captured very well indeed, which validates the use of the string hypothesis in cases when all particles of opposite spins are paired up into dimers.

In order to identify yrast states in the strongly attractive regime it is important to note that breaking a dimer would cost a large amount of energy (${\hbar^2 c^2}/{4m}$). Thus the yrast states of lowest energy for given momentum will be the ones where all fermions are paired, and the pair rapidities $\alpha_m$ are rearranged as permitted by \eref{logaF2b}. The yrast excitations are then obtained by moving a dimer from the inside of the Fermi sphere to the edge and thus creating a hole, as illustrated in \fref{diag_1f} on the right by the filled symbols. Compared to particle-like or mixed particle-hole-like excitations where a dimer rapidity occupies one of the free slots outside the Fermi sphere, the hole excitation always has lower energy at given momentum in direct analogy to the Lieb-Liniger and Tonks-Girardeau models \cite{LL2}. For the rapidities this means omitting one element from the set of ground-state $\ell_m$ quantum numbers of \eref{aF2lms} and adding $\ell_{M+1}$. This operation shifts one of the $\alpha_m$ values and with it, the corresponding pair of charge rapidities with values $\alpha_m \pm i\frac{c}{2}$. 

\begin{figure}[htbp]
\subfigure[]{\includegraphics[width=3.2in]{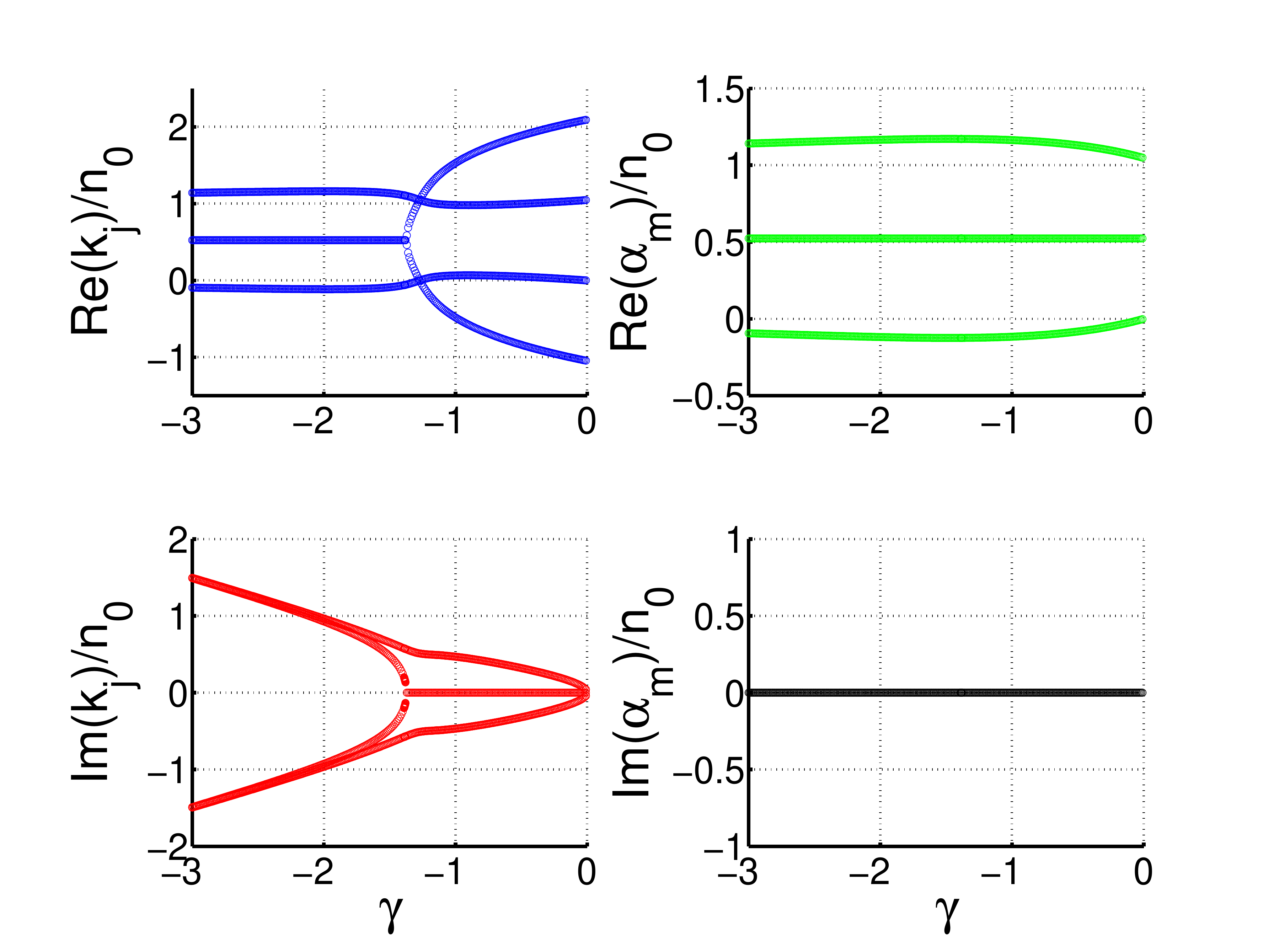}}
\subfigure[]{\includegraphics[width=3.2in]{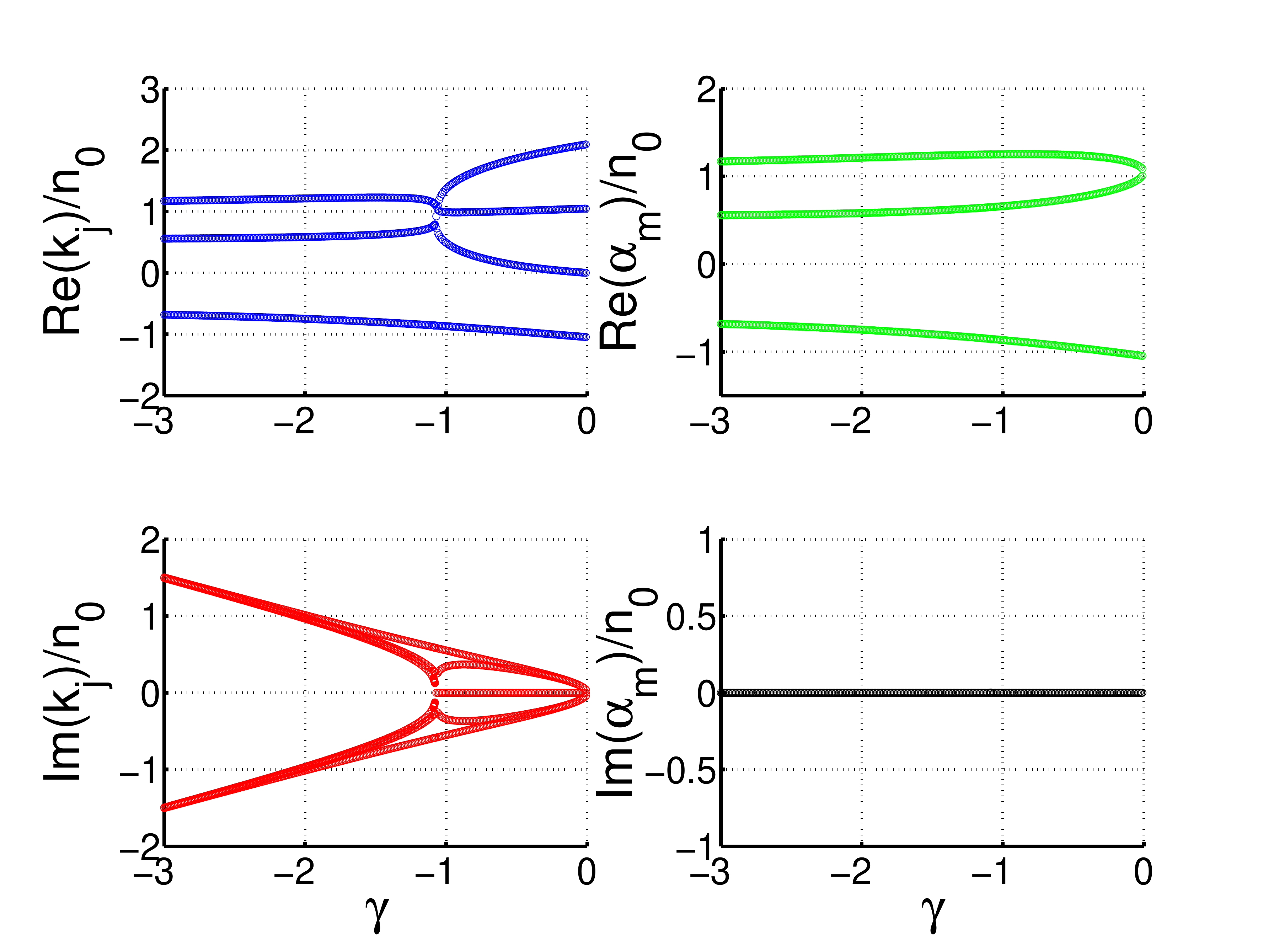}}
\subfigure[]{\includegraphics[width=3.2in]{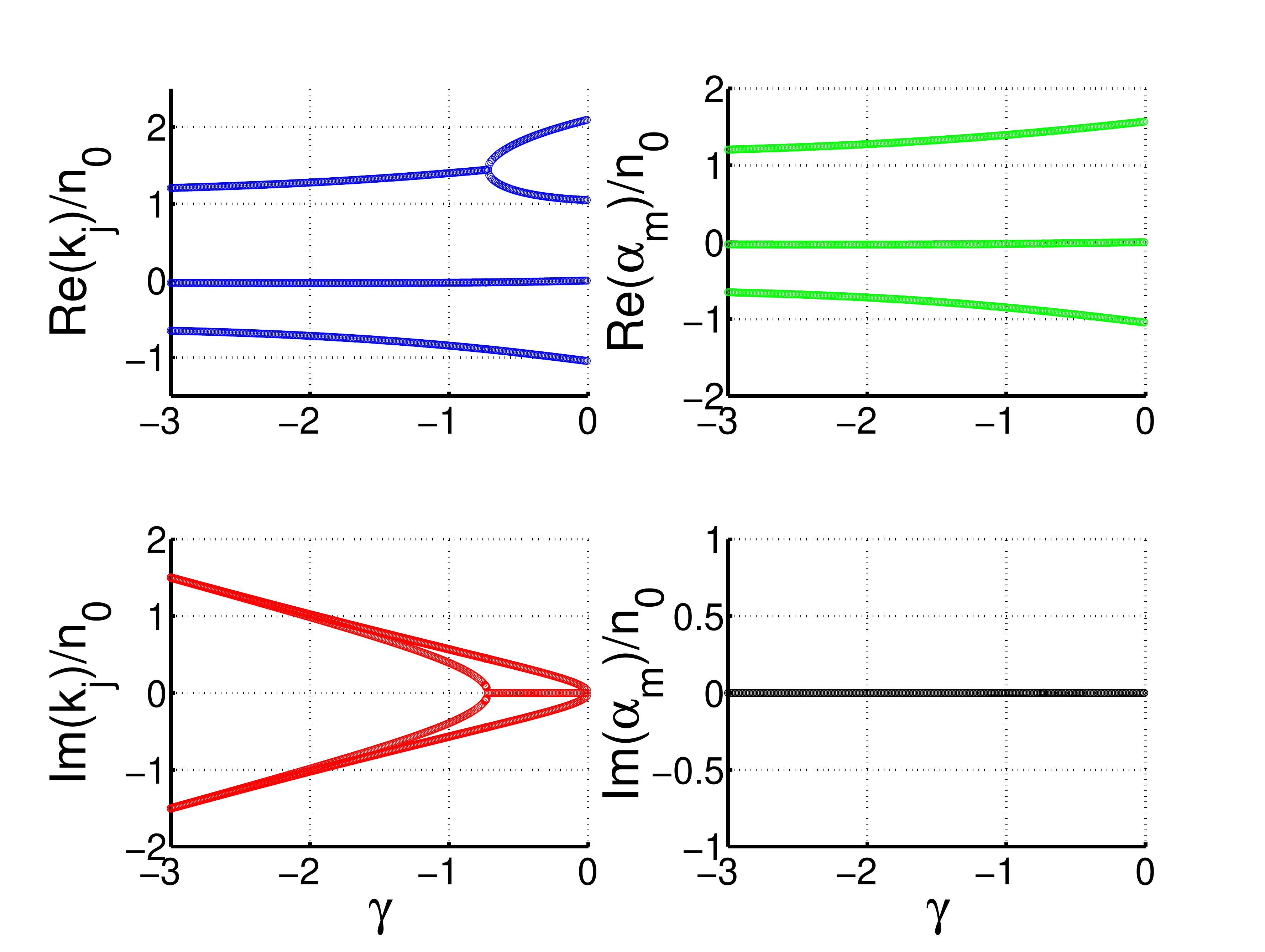}}
\caption{\label{egslns} The real and imaginary parts of charge and spin rapidities for the yrast excitations with $N=6, M=3$. The momentum values for the yrast states are (a) $P_\mathrm{tot} = 3 \hbar \pi/L$, (b) $P_\mathrm{tot} = 2 \hbar \pi/L$, and (c) $P_\mathrm{tot} = 1 \hbar \pi/L$. These solutions were obtained by numerical root finding from the exponential form of the Bethe-ansatz equations  \eref{expF2a} and  \eref{expF2b} starting from the known approximate rapidities for weak interaction.} 
\end{figure}

It is not yet obvious from the asymptotic representations shown in \fref{diag_1f}, whether and how the dimer hole excitations are connected to the single fermion holes we have discussed in the context of weak attractive interactions. For both to be the yrast excitations of the respective regime, it is necessary that they are continuously connected through a change in the interaction parameter. That this is indeed the case is demonstrated in \fref{egslns} where the real and imaginary values of the rapidities are shown as a function of $\gamma$ for $N=6$ particles. Note that the real parts of the charge rapidities converge to the values of the spin rapidities and the imaginary parts become linear in $c$ for strongly negative $c$, as required by the string hypothesis. In addition, we observe that the pair of $k_j$'s that are separated to form the single fermion hole at weak interactions always merge to form a dimer at more negative  values of $c$. Another observation is that the spacing of rapidities on the real axis changes from the single-particle quantization condition $2\pi/L$ at $c=0$ to $\pi/L$ at $c\to -\infty$, which is the appropriate momentum quantization for dimers with mass $2m$.

\begin{figure}[htbp]
\begin{center}
\includegraphics[width=6in]{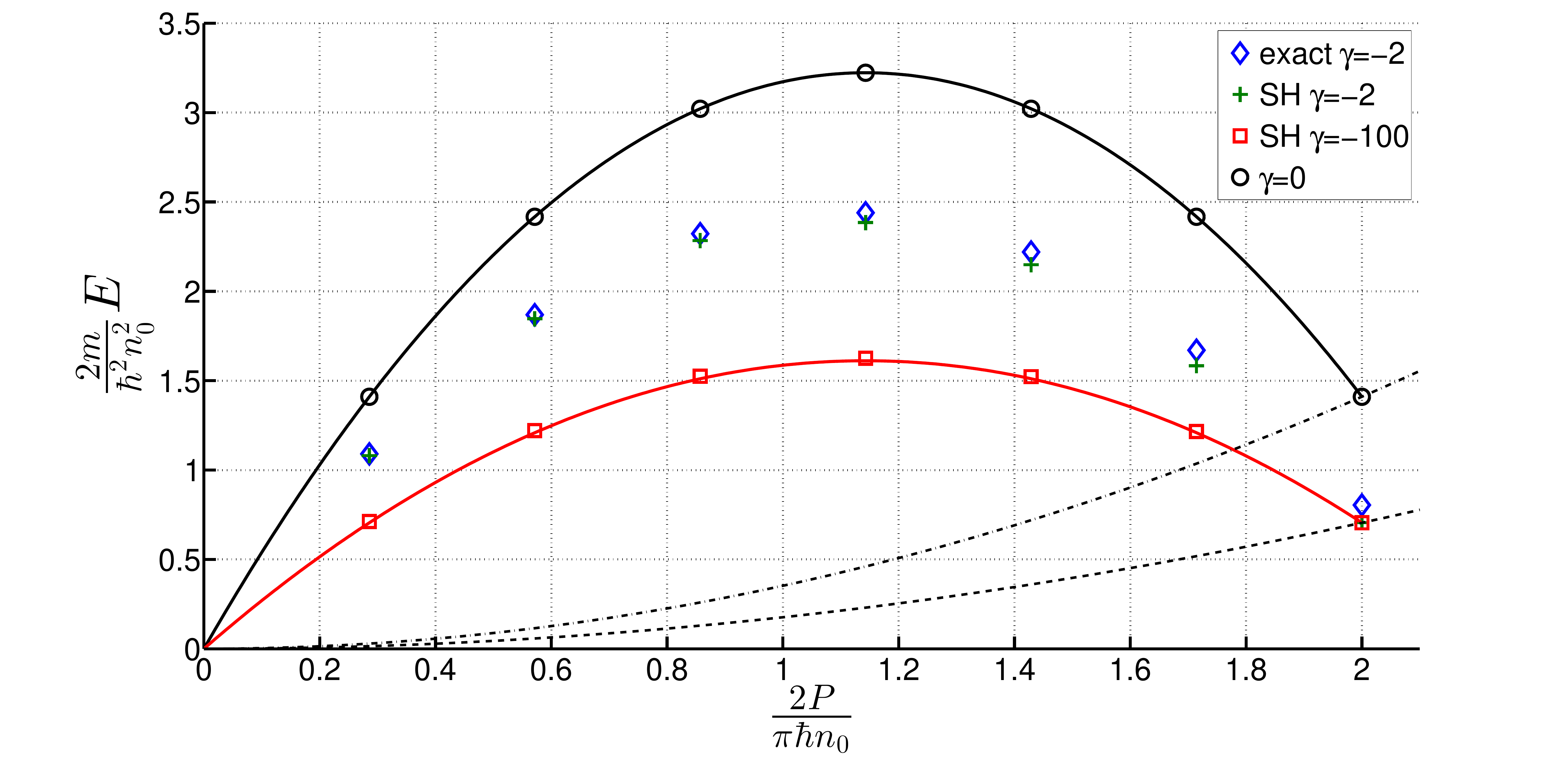}
\caption{\label{single} 
Excitation energy vs.\ momentum for the yrast states shown as symbols for different values of the interaction strength $\gamma$ and for $N=14, M=N/2$. The results for $\gamma = -2$ demonstrate continuity between the exact Bethe-ansatz equations in exponential form \eref{expF2a} and \eref{expF2b} (blue diamonds) and the approximate logarithmic equation \eref{logaF2b}, which is based on the string hypothesis (SH, green crosses). Full lines show the limiting dispersion relations for $\gamma=0$ from \eref{Dapproxw}  in black and for $\gamma \to \infty$ from  \eref{Dapproxs} in red. The dashed line shows the full-system translation parabola $P^2/2mN$ and the dash-dotted line shows the half-system translation parabola $P^2/2mM$.}
\end{center}
\end{figure}

The dispersion relation for the yrast excitation energy $E = E_\mathrm{tot} -E_\mathrm{GS}$ vs.\ momentum is shown in \fref{single}. It is seen that the excitation energy  of each yrast state decreases as the interaction strength changes from noninteracting to strongly attractive within a well defined interval. The limits of the interval are indicated by the full parabolic lines drawn from \eref{Dapproxw} and \eref{Dapproxs} below, respectively. The excitation at momentum $2p_F = \pi \hbar n_0= M\times 2\pi \hbar/L$ deserves special attention. This value is half of the umklapp momentum  $4p_F$ discussed in \sref{YGinteracting} and is also the  momentum value obtained by adding the unit value $2\pi /L$ to every single-particle wavenumber of half the fermions, e.g.\ to only the spin-up fermions. In the noninteracting case, the corresponding yrast excitation is  obtained by taking a spin-up fermion from the left side of the Fermi sphere and attaching it to the right side, which is equivalent to  shifting the Fermi sphere by one unit wavenumber while the spin-down Fermi sphere remains fixed. As can be expected, the energy of this excitation (for $\gamma=0$) lies on the parabola $P^2/2mM$, which we refer to as the half-system translation parabola. It is shown as a dash-dotted line in \fref{single}. For finite attractive interaction strength $\gamma<0$, the energy of this excitation decreases and eventually moves to the full system translation parabola $P^2/2mN$, which is shown as a dashed line. This is to be expected, since we know that the system dimerizes for $\gamma \to -\infty$, and an umklapp excitation of a Tonks-Girardeau gas of $M=N/2$ dimers can be obtained at momentum $2p_F$ with energy $P^2/(2\times 2m\times M)$.

Since we know the ground- and excited-state rapidities in both limits, it is possible to find the asymptotic dispersion relations for weak and strong interactions
\begin{eqnarray}
\label{Dapproxw}
E^\mathrm{weak} &=& \frac{4p_F^2}{2m}\left[-\left(\frac{P}{2p_F}\right)^2+\frac{P}{2p_F}\left(1+\frac{2}{N}\right)-\frac{2\gamma}{\pi^2N}\right],\\
\label{Dapproxs}
E^\mathrm{strong} &=& \frac{p_F^2}{2m}\left[-\frac{1}{2}\left(\frac{P}{p_F}\right)^2+\frac{P}{p_F}\left(1+\frac{2}{N}\right)\right],
\end{eqnarray}
where $p_F=\pi\hbar n_0/2$ is the Fermi momentum of the free system. These dispersions are added to figure \ref{single} as solid lines.

Moreover, Fig.~\ref{single} compares the exact and string hypothesis dispersions at $\gamma=-2$, roughly the lowest $c$ value that can be tackled by the exact exponential equations at the given number of particles. The small difference visible between the dispersions is due to the fact that we have $M$ complex $k_j$ pairs, which directly contribute to the energy and their imaginary parts in the exact and approximate equations are different.
\subsection{\label{sTG}Connection to the super Tonks-Girardeau gas}
The gaseous phase of attractive bosons known as the super Tonks-Girardeau gas can be described by the regular Bethe ansatz equations for the Lieb-Liniger model of bosons in 1D under the additional restriction that all rapidities are real valued such that bound states of bosons are excluded \cite{Batchelor2005}. These equations map one-to-one to the equations of the attractive Yang-Gaudin model \eref{expF2a} and \eref{expF2b} as well as \eref{logaF2b} under the string hypothesis \cite{Batchelor2010}, if we identify $M=N/2$ as the number of bosons (dimers) with mass $m_b = 2 m$, scattering length $a_b=a_{1D}/2$, $c_b= 2c$, and dimensionless coupling parameter $\gamma_b \equiv c_b/n_b= 4 \gamma$ with $n_b=n_0/2$. Thus all results of this work pertaining to the dimerized attractive spin-$\frac{1}{2}$ Fermi gas apply to the super Tonks-Girardeau gas as well. An important physical difference between the models is that  bound states of fermion dimers are excluded by the Pauli exclusion principle \cite{Mora2005}, while bound states of bosons exist in a lower energy sector. The yrast states of the spin-$\frac{1}{2}$ Fermi superfluid are thus ground states of their respective momentum sector while the corresponding super Tonks-Girardeau yrast states are highly excited.
\section{\label{TDL_eqns}Yrast dispersion in the thermodynamic limit}
In the thermodynamic limit where $N,L\rightarrow\infty$ while $n_0 = N/L$ remains fixed, the unit of momentum discretization $ 2\pi \hbar/L$ becomes infinitesimally small while the full- and half-system umklapp momenta, $4p_F$ and $2p_F$, remain finite and meaningful quantities with $p_F = \pi \hbar n_0/2$. The yrast dispersion is therefore replaced by a continuous function and the sets of rapidities are replaced by continuous distributions. The interaction strength, $\gamma = \frac{c}{n_0}$, and the ratio $M/N$ (kept at a value of $\frac{1}{2}$ for the purpose of this work), are the only two dimensionless parameters characterizing the ground state, while the yrast states, in addition, are determined by the momentum $P/p_F$.

In the finite-system case we saw that the string hypothesis works when $|c|L\gg1$ and the exact Bethe ansatz equations can be replaced by the logarithmic form \eref{logaF2b} for fully dimerized states. In the thermodynamic limit, this condition is fulfilled for any value of the interaction strength $\gamma$ since $L\to\infty$, and thus we only need to consider the logarithmic equations. These coupled algebraic equations become Fredholm integral equations of the second kind, first derived by Gaudin \cite{GaudinThesis}. The ground state integral equations can be found in  \cite{Guan2011} and there is some earlier work examining excitations in the Yang-Gaudin model \cite{GaudinThesis,Krivnov,Schlottmann,Ma2012}, often focusing on spin-changing excitations. In the following we present the integral equations that are appropriate for the ground state and yrast excitations. These are solved numerically using the  \texttt{Fie} package for MATLAB \cite{Atkinson2008}.
\subsection{\label{TDL_aF2gs}Ground state}
\begin{figure}[htbp]
\begin{center}
\includegraphics[width=6in]{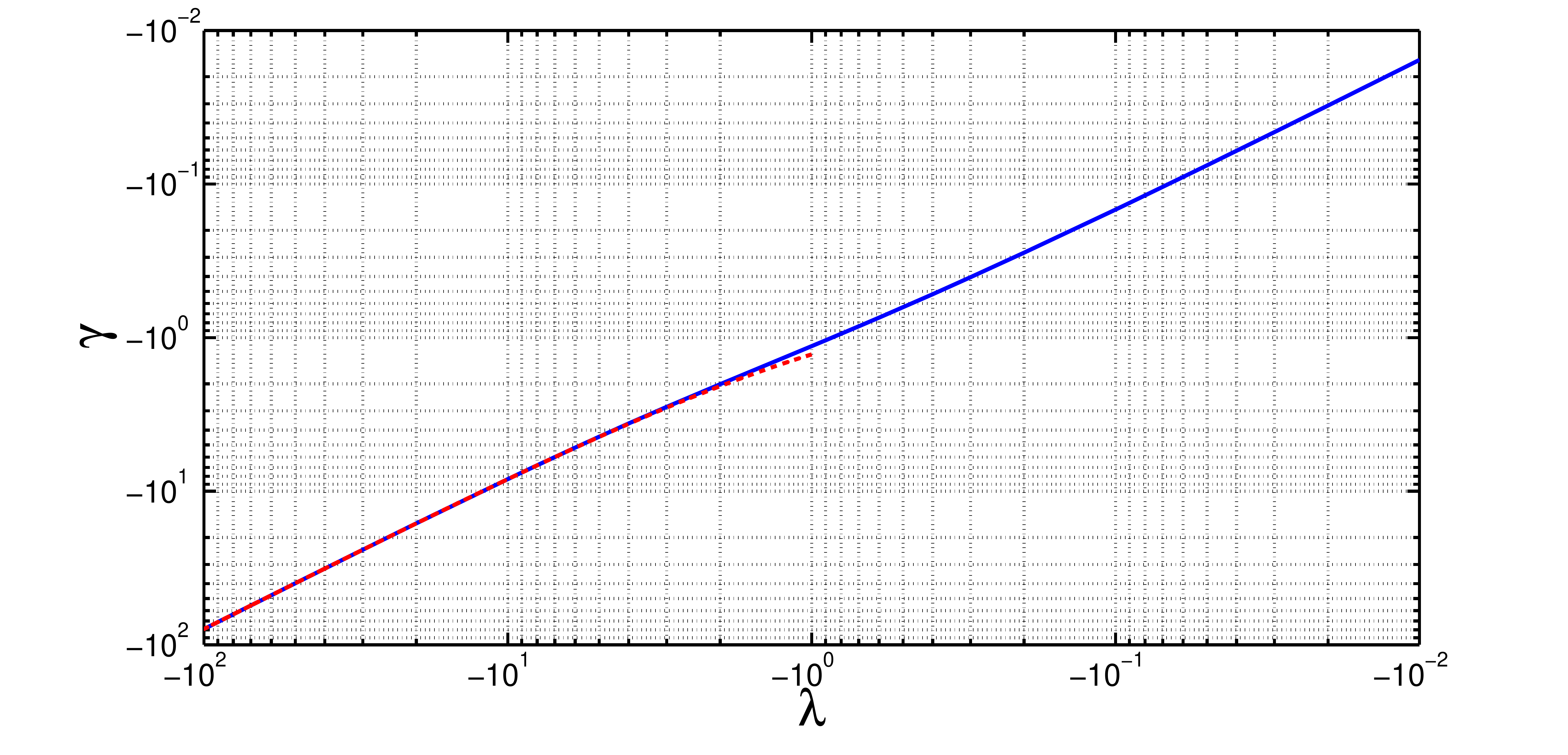}
\caption{\label{lamgam_aF2} The dependence of the physical dimensionless interaction strength $\gamma= c/n_0$ on the computational parameter $\lambda$ (full line). The dashed line shows the asymptotic relation \eref{eq:gammaapprox} for the strongly attractive regime.}
\end{center}
\end{figure}

The balanced (fully-dimerized) ground state properties can be found by introducing the density distribution function $r(\alpha)$ for the dimer rapidities, and rewriting  \eref{logaF2b} into the integral equation \cite{GaudinThesis,Guan2011}
\begin{equation}
\frac{1}{\pi} = r(\alpha) - \frac{c}{\pi}\int\limits_{-b}^{b}\frac{r(\alpha')}{c^2+(\alpha-\alpha')^2}d\alpha' ,
\end{equation}
with the constraint 
\begin{equation}
\int\limits_{-b}^{b}r(\alpha)d\alpha = \frac{M}{L},
\end{equation}
where $b$ is the boundary of the rapidity distribution. We will follow the common convention \cite{Guan2011} to  redefine the reference point for energy measurements by subtracting the binding energy $-\frac{\hbar^2}{2m}Mc^2/2$, which diverges in the strongly attractive limit. This does not affect the dispersion relation, which reports the energy difference between excited and ground state. The ground state energy and momentum then become
\begin{eqnarray}
E_\mathrm{GS} &=& \frac{\hbar^2}{2m}L\int\limits_{-b}^{b}2\alpha^2r(\alpha)d\alpha,\\
P_\mathrm{GS} &=& \hbar2L\int\limits_{-b}^{b}\alpha r(\alpha)d\alpha = 0.
\end{eqnarray}
In order to solve the integral equation numerically we scale the equations, introducing the dimensionless rapidity $x$ and the parameter $\lambda$:
\begin{equation}
\label{aF2scale1}
x= \frac{\alpha}{b}, \quad \lambda = \frac{c}{b} = \frac{n_0 \gamma}{b},\quad g(x)\equiv r(bx)  .
\end{equation}
We thus obtain
\begin{equation}
1 = \pi g(x) - \lambda\int\limits_{-1}^{1}\frac{g(x')}{\lambda^2+(x-x')^2}dx',
\end{equation}
with the normalization condition
\begin{equation} \label{eq:lamgam}
\lambda = 2\gamma\int\limits_{-1}^{1}g(x)dx,
\end{equation}
where $M=N/2$ has been used.
The parameter $\lambda$ has a monotonic relationship to the dimensionless interaction strength $\gamma$, as is shown in \fref{lamgam_aF2}. For numerical calculations it is practical to set $\lambda$ first and then obtain the value of $\gamma$ from \eref{eq:lamgam}.

The ground state energy now can be written as
\begin{equation}
\label{edef}
E_\mathrm{GS} = N \frac{\hbar^2 n_0^2}{2m} e(\gamma),
\end{equation}
where $e(\gamma)$ is defined by 
\begin{equation}
\label{e}
e(\gamma) = 2\left(\frac{\gamma}{\lambda}\right)^3\int\limits_{-1}^{1}x^2g(x)dx.
\end{equation}
Knowledge of the the ground state energy provides us with access to the chemical potential 
\begin{equation}
\label{mu}
\mu \equiv \frac{\rmd E_\mathrm{GS}}{\rmd N}= \frac{\hbar^2}{2m}n_0^2\left(3e - \gamma\frac{\rmd e}{\rmd \gamma}\right) ,
\end{equation}
and the speed of sound obtained through the compressibility \cite{LL2}
\begin{align}
\label{speedofsound}
v_c = \sqrt{\frac{n_0}{m} \frac{\rmd \mu}{\rmd n_0}}.
\end{align}
Note that the chemical potential as defined here does not contain the contribution $-\hbar^2 c^2/4m$ from the binding energy of the dimer that has to be broken in order to remove a single fermion. The chemical potential for the (bosonic) dimers is $2 \mu$.  
\subsection{\label{TDL_aF2s}Yrast dispersion relation}
\begin{figure}[htbp]
\begin{center}
\includegraphics[width=6in]{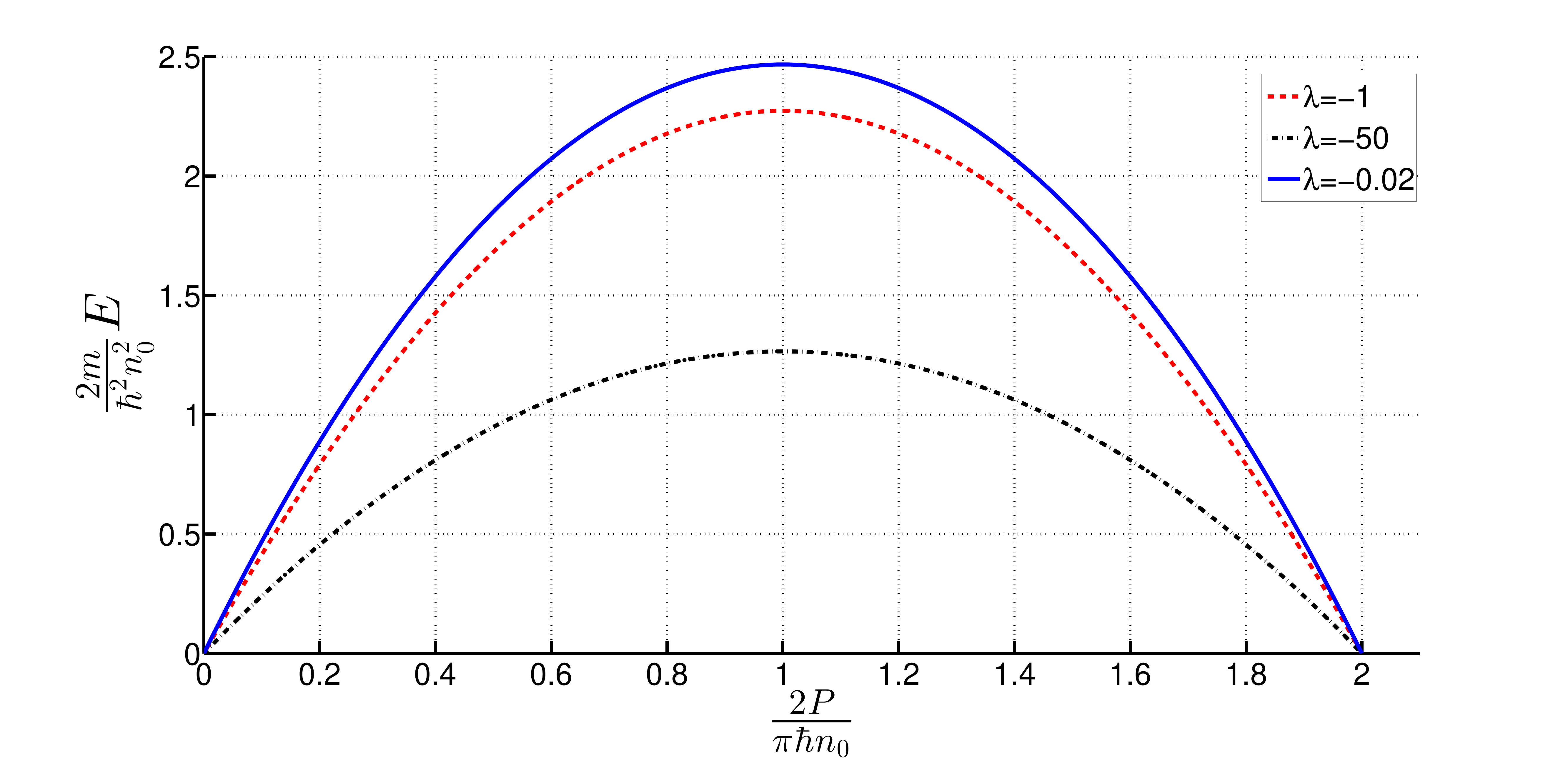}
\caption{\label{daF2_single} Yrast dispersion relation (dimer hole excitations) in the thermodynamic limit from \eref{eq:Ptdl} and \eref{eq:Etdl}. Blue solid line: $\lambda=-0.02$ ($\gamma\approx -0.031$), red dashed line: $\lambda=-1$ ($\gamma \approx -1.1$), black dash-dotted line: $\lambda=-50$ ($\gamma \approx-40$). The yrast dispersion relation is periodic in momentum with period $2p_F=\pi \hbar n_0$. Only one period is shown. For the whole range of interactions, the dispersion relation relates to a negative-effective-mass quasiparticle and is reminiscent of dark solitons in the nonlinear Schr\"odinger equation \cite{pitaevskii03:book}.}
\end{center}
\end{figure}

The yrast excited states are characterized by a single hole in the set of dimer rapidities but the rapidities are shifted from their ground state positions. An integral equation can then be derived for a function $J(\alpha)$, which quantifies the shift \cite{LL2,GaudinThesis}. \ref{TDLhowto} provides some further details on this derivation. The  integral equation has a similar structure to that of the ground state but contains the hole rapidity $Q$ as an additional parameter with $-b < Q < b$:
\begin{equation}
2\pi J(\alpha) = 2c\int\limits_{-b}^{b}\frac{J(\alpha')}{c^2+(\alpha-\alpha')^2}d\alpha' + \pi-\theta\left(\alpha-Q\right).
\end{equation}
The energy and momentum of the excitation relative to the ground state are found from
\begin{eqnarray}
P &=& \hbar\left[-Q + 2\int\limits_{-b}^{b}J(\alpha)d\alpha\right],\\
E &=& \frac{\hbar^2}{2m}\left[-2Q^2 + 4\int\limits_{-b}^{b}\alpha J(\alpha)d\alpha\right] + 2\mu.\ \ \ \ 
\end{eqnarray}
The equations can be made dimensionless by a similar rescaling as for the ground state with the additional dimensionless quantities
\begin{equation}
\label{aF2scale2}
q = \frac{Q}{b},\quad  h(x) \equiv J(bx) .
\end{equation}
The integral equation becomes
\begin{equation}
\pi h(x) = \lambda\int\limits_{-1}^{1}\frac{h(x')}{\lambda^2+(x-x')^2}dx' + \frac{\pi}{2} -  \tan^{-1}\left(\frac{q-x}{\lambda}\right),
\end{equation}
and the excitation energy and momentum are obtained from the dimensionless function $h(x)$ by
\begin{eqnarray} \label{eq:Ptdl}
P &=& \hbar 2b\left[- q + \int\limits_{-1}^{1}h(x)dx\right],\\ \label{eq:Etdl}
E &=& \frac{\hbar^2}{2m} 2b^2\left[-q^2 + 2\int\limits_{-1}^{1}xh(x)dx\right]+ 2\mu.
\end{eqnarray}
Figure \ref{daF2_single} shows example dispersions for this excitation branch. Limiting expressions for weak and strong interactions are easily obtained from the finite particle number results \eref{Dapproxw} and \eref{Dapproxs}, respectively, by taking the limit $N \to \infty$. Note that the translation parabolas $\propto P^2/Nm$  move to the $P$-axis in the thermodynamic limit as $N\to\infty$. Thus both the umklapp and half-umklapp excitations have zero energy and the dispersion relation $E(\mu,P)$ is periodic in momentum with period $2p_F$.

The yrast dispersion relation is gapless and, as demanded from general principles, the slope at the origin is the speed of sound
\begin{align}
 \lim_{P\to +0} \frac{\partial E(\mu,P)}{\partial P} = v_c ,
\end{align}
as in the Lieb-Liniger model \cite{LL2}. A dispersion relation for phonon-like ``particle'' excitations with the same slope at the origin but higher energy than the yrast excitations can also be obtained from the Bethe ansatz \cite{GaudinThesis}. In addition there is a triplet of elementary excitations that involve breaking a single dimer pair. The corresponding dispersion relations are gapped for $\gamma<0$ and also have higher energy than the yrast excitations.

An explicit form of the dispersion relation can be obtained in the strongly attractive regime by solving the integral equations to leading order in $\lambda^{-1}$. Neglecting $(x-x')^2$ compared to $\lambda^2$ in the kernel, we find $g=(\pi-2/\lambda)^{-1}$, a constant independent of $x$. Expanding the inhomogeneous term in the excited-state equation to first order about zero, naturally leads to a linear ansatz for $h(x)$: $h(x) = \alpha(\lambda,q) + \beta(\lambda,q)x$. Indeed, $\alpha(\lambda,q)=\frac{\pi\lambda-2q}{2(\pi\lambda-2)}$ and $\beta(\lambda)=\frac{1}{\pi\lambda}$ satisfy the resulting equation. We can now evaluate all the ground state properties:
\begin{eqnarray}
b &=& n_0\frac{\pi\lambda-2}{4\lambda},\\\label{eq:gammaapprox}
\gamma &=& \frac{1}{2}\left(\frac{\pi\lambda}{2}-1\right),\\
e &=& \frac{1}{48}\left(\pi-\frac{2}{\lambda}\right)^2,\\
\mu &=& \frac{\hbar^2 n_0^2}{2m} \left[\frac{\pi^2}{16}-\frac{\pi}{3\lambda}\right].
\end{eqnarray}
The excitation energy and momentum can also be obtained, giving
\begin{eqnarray}
P &=& \hbar 2b\left[-q+2\alpha\right],\\
E &=& \frac{\hbar^2}{2m}2b^2\left[-q^2+\frac{4\beta}{3}\right]+2\mu.
\end{eqnarray}
Eliminating $q$ and substituting for $\lambda$ in terms of $\gamma$, the dispersion relation becomes
\begin{equation}
E = \frac{p_F^2}{2m}\left(\frac{P}{p_F}- \frac{P^2}{2p_F^2}\right) \left(\frac{2\gamma}{1+2\gamma}\right)^2 + \Or\left(\gamma^{-3}\right).
\end{equation}
This result agrees with \eref{Dapproxs} for $\gamma\to-\infty$ and $N\to\infty$, as it should.
\subsection{\label{Ns}Missing particle number and phase step}
The dispersion relations discussed in the previous section give us access to a number of quantities that define the character of soliton-like excitations, and that can be derived assuming the existence of a localized quasiparticle. In the following we will consider $E=E(\mu,P)$ as a function of the two variables $\mu$ and $P$. Note that both the density $n_0$ and $\gamma=c/n_0$ depend on the chemical potential $\mu$ while parameters of the Hamiltonian, in particular $c$, do not. 

Associating the yrast excitations with a density dip in the inhomogeneous density $n_s(x)$ with background density $n_0$ allows us to define the particle number  $N_d = \int (n_s - n_0) \rmd x$. The quantity $N_d$ can then be obtained from thermodynamic relations \cite[equation (5)]{Schecter2012} and written as
\begin{equation}
\label{Schectdef}
N_d = - \left({\frac{\partial E}{\partial \mu} + \frac{v_s}{n_0}\frac{\rmd n_0}{\rmd \mu}P}\right) \left({1-\frac{mv_s^2}{n_0}\frac{\rmd n_0}{\rmd \mu}}\right)^{-1}.
\end{equation}
An alternative derivation of \eref{Schectdef} based on  the Hellman-Feynman theorem in a co-moving reference frame will be published elsewhere \cite{Sophie:unpup}. The quantity $N_d$ differs in general  from $N_s\equiv m_P/m =-\rmd E/\rmd \mu|_{v_s}$ of \cite{Scott2011,Astrakharchik2012} but gives identical results for $v_s=0$ as can be seen directly from \eref{Schectdef}.  For dark solitons as well as for the yrast excitations considered here, $N_d<0$ and thus is also referred to as the missing particle number, i.e.\ the number of particles that make up the difference between the soliton's density profile and a constant background density. Strictly speaking, we are using the term only by analogy to classical solitons, e.g.\ in the nonlinear Schr\"odinger equation, since the physical meaning of $N_d$ for yrast excitations in general is not fully understood. Likewise it is possible to assign to the states on the yrast dispersion a phase step $\Delta \phi$. As a well-known characteristic of dark solitons in the nonlinear Schr\"odinger equation \cite{Tsuzuki1971}, the phase step becomes relevant in a system with periodic boundary conditions, where it determines the size of the backflow current: the entire superfluid flows at $v_{cf}=\hbar\Delta\phi/(2mL)$. This counter-flow  is induced in order to counteract the phase jump of the soliton and to ensure that the total phase is well defined and continuous across the periodic boundary. $\Delta\phi$ can  be calculated from \cite{Scott2011}
\begin{align}
P = P_s + \frac{1}{2} \hbar n_0 \Delta\phi,
\end{align}
which states that the total (or canonical) momentum $P$ in a periodic box is the sum of $P_s=mv_sN_d$, the physical momentum of the soliton, and the counter-flow momentum of the entire superfluid. The phase step $\Delta \phi$ is thus well defined, even though 1D quantum gases are thought to possess at most  a fluctuating condensate phase \cite{petrov00}.

\begin{figure}[htbp]
\subfigure[]{\includegraphics[width=3.2in]{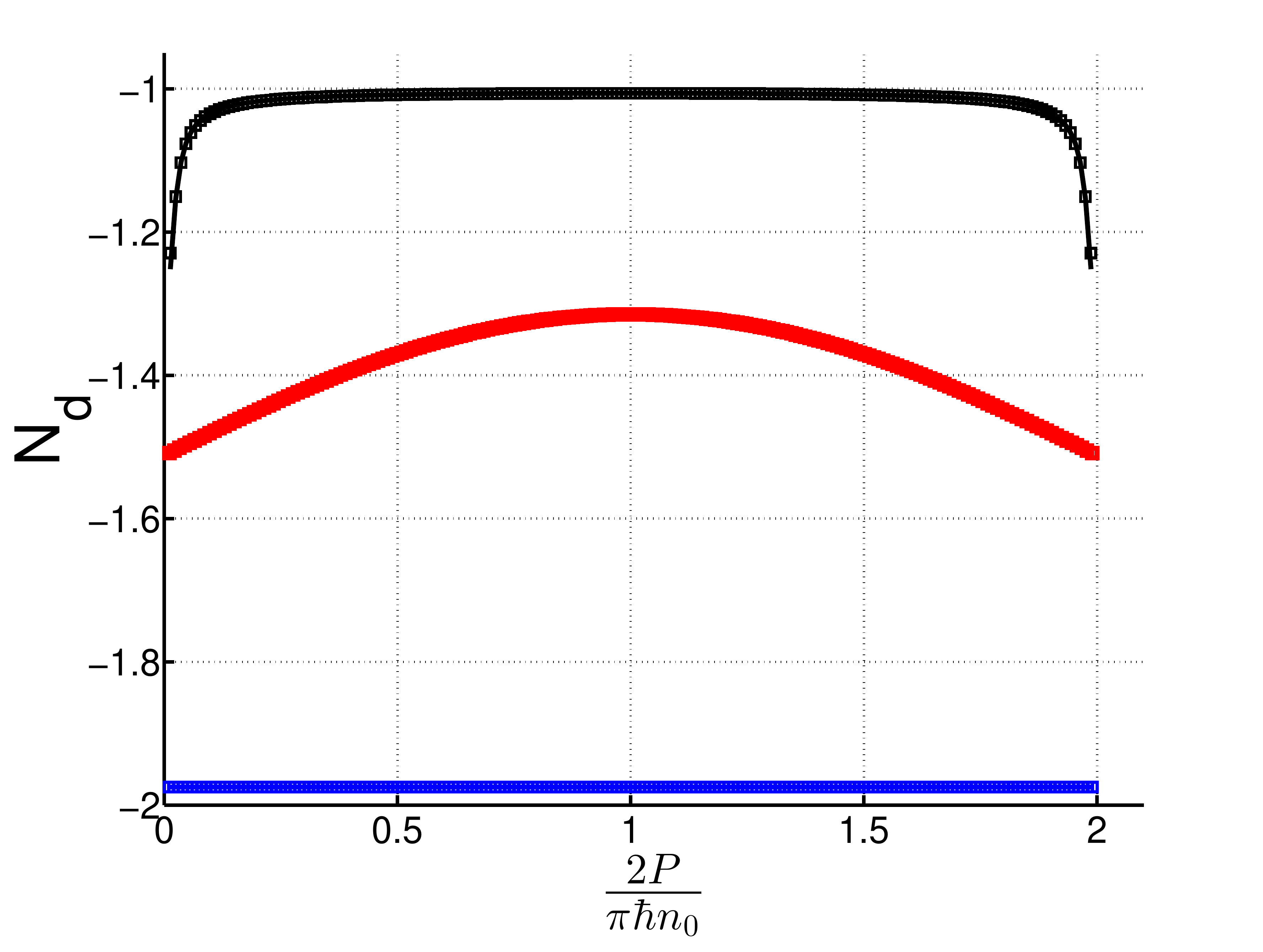}}
\subfigure[]{\includegraphics[width=3.2in]{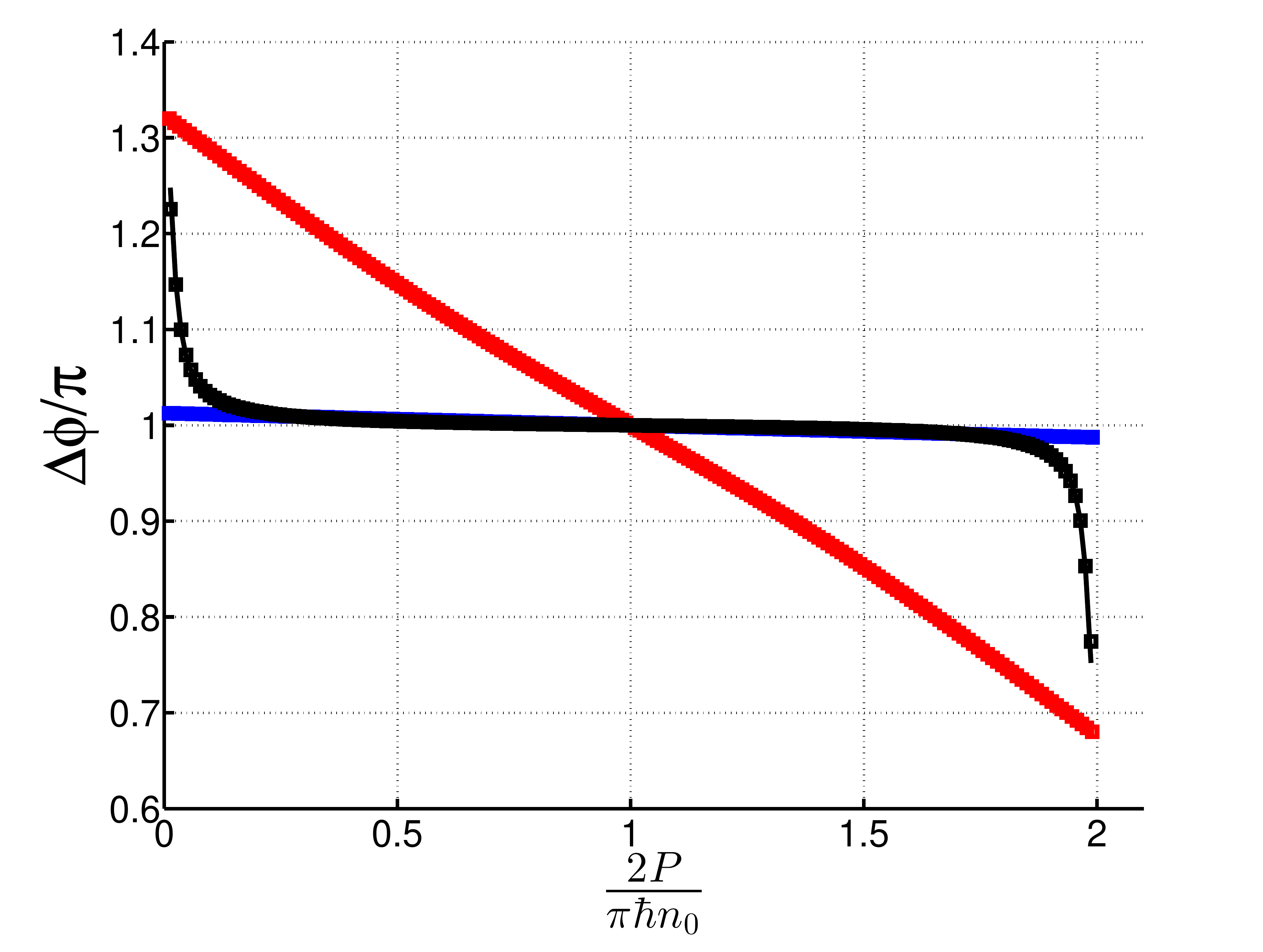}}
\caption{\label{NsDpaF2_single} Missing particle number (a) and phase difference (b) for single fermion holes in the $\gamma\rightarrow0$ limit, which become single dimer holes in the $\gamma\rightarrow -\infty$ limit. Black: $\lambda=-0.02, \gamma=-0.030815$, red: $\lambda=-1, \gamma=-1.1363$, blue: $\lambda=-50, \gamma=-39.770$. Direct evaluation of (\ref{Schectdef}) yields the results plotted as squares, while those from Campbell's formulas \eref{CampNs} and \eref{CampDp} are shown as solid lines.}
\end{figure}

The missing particle number $N_d$ and the phase step $\Delta \phi$ along the dispersion relation are shown for different values of the interaction strength in figure \ref{NsDpaF2_single}. Details about evaluating (\ref{Schectdef}) are presented in \ref{Ns_aF2s}. The limiting cases of weak and strong interactions are easy to interpret, as $N_d$ becomes constant as a function of momentum and assumes the values -2 for a dimer hole in the Tonks-Girardeau gas of dimers ($\gamma \to -\infty$) and -1 for a single fermion hole in the non-interacting limit ($\gamma = 0$). Also the phase step assumes the constant value of $\pi$ in both limits. A most interesting behaviour is found in the crossover regime, where both $N_d$ and $\Delta \phi$ show significant momentum dependence.

For the Lieb-Liniger model, A.~Campbell \cite{Campbell2013} derived convenient equations for $N_d$ and $\Delta\phi$ that avoid numerical derivatives completely and are thus very efficient to evaluate (see just under equation (5.68) of \cite{Campbell2013}). With trivial adaption to the current case of the Yang-Gaudin model they read \footnote{The constant -2 in \eref{CampNs}  was missing in \cite{Campbell2013} but is needed to achieve consistency with \eref{Schectdef}. For the Lieb-Liniger model it should be replaced by -1 and $\Delta\phi$ should have an additional factor of 1/2.}:
\begin{eqnarray}
\label{CampNs}
N_d &=& -2\pi g(1)\left[h(1)-h(-1)\right]-2,\\
\label{CampDp}
\Delta\phi &=& \frac{1}{g(1)}\left[h(1)+h(-1)\right],
\end{eqnarray}
where the functions $g(x)$ and $h(x)$ were introduced in sections \ref{TDL_aF2gs} and \ref{TDL_aF2s} and still depend on the interaction parameter $\gamma$ (and the momentum in the case of $h(x)$). The excellent agreement with the direct evaluation of \eref{Schectdef} is seen by the overlapping curves in figure \ref{NsDpaF2_single}.

In the strongly-interacting limit, using the approximate analytical solutions of the Bethe ansatz integral equations and Campbell's formulae, we find the following limiting expressions: 
\begin{align}
N_d&=-2-\frac{1}{\gamma} +\Or\left(\gamma^{-2}\right) ,\\
\Delta\phi &= \pi \left[1 + \left(\frac{P}{p_F}-1\right)(1+2\gamma)^{-1} \right] + \Or\left(\gamma^{-2}\right).
\end{align}
\subsection{\label{Masses}Quasiparticle dynamics in a harmonic trap}
Konotop and Pitaevskii have argued that the motion of solitons on a slowly varying background is governed by the principles of Landau's quasiparticle dynamics, which leads to simple classical equations of motion \cite{Konotop2004}. Here we apply these ideas to yrast excitations in the inhomogeneous 1D Fermi gas. The crucial point is to assume that localized quasiparticles can be formed from the yrast excitations, e.g.\ by wave-packet formation, which move with a velocity $v_s = \partial E(\mu,P)/\partial P$. There is good reason to expect this to happen given the evidence connecting yrast excitations to dark solitons in the closely related Lieb-Liniger model as discussed in the introduction.

Suppose now the background variations happen on a length scale $R_\mathrm{TF}$ that is large compared to the quasiparticle length scale $\ell_s$ and  the Fermi wavelength $\xi_F = 4/n_0$. Since the quasiparticle velocity $v_s$ is bounded by the speed of sound $v_c \sim v_F = \pi n \hbar/2m$, we can also suppose that the acceleration of the quasiparticle moving over the slowly varying background is small enough to suppress the radiation of energy. In the local density approximation, the chemical potential $\mu(X)= \mu_0 - V(X)$ varies locally with position and adjusts to the value of the external potential. Requiring the energy of the quasiparticle $E=E(\mu,P)$ to be conserved during the motion then leads to a Hamiltonian equation of motion for the position $X(t)$ of the quasiparticle
\begin{align}
0 =& - \frac{m_P}{m} \frac{\rmd \mu}{\rmd X} \dot{X} + v_s m_I  \frac{\rmd v_s}{\rmd t} 
\end{align}
where we have identified the inertial mass  
\begin{equation}
m_I= \left. \frac{\rmd P}{\rmd v_s}\right|_\mu = \left(\frac{\partial^2 E}{\partial P^2}\right)^{-1}  
= \frac{1}{v_s} \left.\frac{\rmd E}{\rmd v_s}\right|_\mu,
\end{equation} 
and the physical mass \cite{Mateo2015a} (see also \cite{Scott2011,Liao11pr:FermiSolitons,Yefsah})
\begin{align}
 m_P = - m\left( \frac{\partial E}{\partial \mu} - m_I\, v_s \frac{\partial^2 E}{\partial \mu \,\partial P}\right) = -m \left. \frac{\rmd E}{\rmd \mu}\right|_{v_s} .
\end{align}
At zero velocity it is closely related to the missing particle number
\begin{align}
m_P = m N_d \quad \textrm{at} \; v_s = 0, 
\end{align}
as can be seen from \eref{Schectdef}. It takes on negative values and plays the role of a buoyancy parameter, or bubble mass, as it determines the value of the buoyancy force 
\begin{align}
\frac{m_P}{m} \frac{\rmd \mu}{\rmd X} = -\frac{m_P}{m}  \frac{\rmd V}{\rmd X}
\end{align}
Assuming a harmonic trapping potential $V(x)=\frac{1}{2}m\omega_\mathrm{trap}^2 x^2$, the quasiparticle equation of motion finally takes the form
\begin{align} \label{eom}
m_I \ddot{X} + m_P\, \omega_\mathrm{trap}^2 X = 0 .
\end{align}
Since both mass parameters are negative and approximately constant at small velocities, \eref{eom} is solved by harmonic oscillations with a frequency $\Omega$ given by \eref{eq:massratio}.

\begin{figure}[htbp]
\begin{center}
\includegraphics[width=6in]{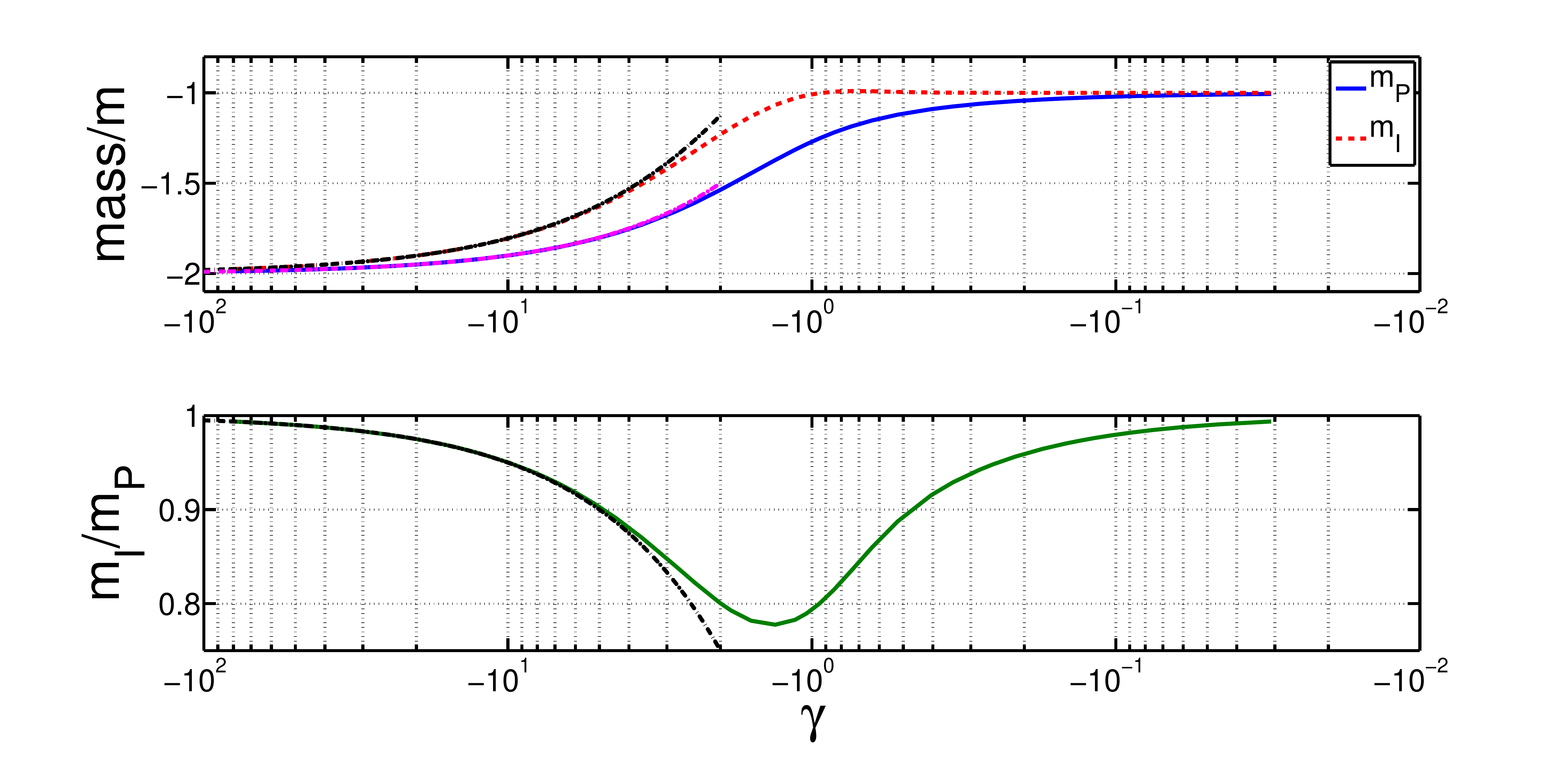}
\caption{\label{maF2_single} Physical and inertial masses (a) and their ratio (b) for yrast excitations of the balanced spin-$\frac{1}{2}$ Fermi gas with attractive interactions. The dash-dotted lines are the approximate equations, top panel: (\ref{approxm_P}) in magenta, (\ref{approxm_I}) in black and bottom panel: (\ref{approxratio}) in black. These are in excellent agreement with the exact results for $\gamma \lesssim-5$.}
\end{center}
\end{figure}

The inertial and the physical masses at $v_s=0$ can be calculated from the dispersion relation and are shown in figure \ref{maF2_single} along with their ratio, $m_I/m_P$. The limiting value of 1 for $m_I/m_P$ in the weakly and strongly interacting regimes means that quasiparticles will oscillate in the harmonically trapped gas with the same frequency as a single fermion would, or, indeed, as the centre-of-mass coordinate of an interacting cloud would if dipole oscillations were initiated \footnote{Due to the separability of centre-of-mass and relative motion in a harmonic trapping potential, this is true regardless of the strength of the inter-particle interaction.}. This is not unexpected, since both limits correspond to that of non-interacting fermions: $\gamma =0$ is a free gas of spin-$\frac{1}{2}$ fermions and $\gamma \to -\infty$ is a Tonks-Girardeau gas of dimers, which maps to a gas of spinless fermions by the Bose-Fermi mapping \cite{girardeau60}. The yrast excitations become free single particle holes in these limits, single fermion holes for $\gamma=0$ and single dimer holes for  $\gamma \to -\infty$, and thus should oscillate with the trap frequency. 

In the strongly attractive regime we can obtain expressions for the masses and their ratio from the analytical solutions of the integral equations in the 
limit $\gamma\rightarrow-\infty$:
\begin{eqnarray}
\label{approxm_P}
m_P &=&-2m \frac{2\gamma+1}{2\gamma} ,\\
\label{approxm_I}
m_I &=& -2m\left(\frac{2\gamma+1}{2\gamma}\right)^2,\\
\label{approxratio}
\frac{m_I}{m_P} &=& \frac{2\gamma+1}{2\gamma}.
\end{eqnarray}
These are shown in figure \ref{maF2_single} as dash-dotted lines and good agreement with numerics is found for $\gamma \lesssim -5$. It is seen that for finite $\gamma$ the mass parameters depart from the free dimer value of $-2m$, which is reminiscent of polarons being dressed by phonon excitations. The value of such an analogy is limited, however as the holes get lighter as $\gamma$ increases, and the quasiparticle is a bosonic excitation. For the super Tonks-Girardeau gas in particular this means that the hole-like excitations become lighter than the bare boson mass $m_b=2m$. Note that the strongly attractive regime is the one where the super Tonks-Girardeau gas has the longest life time \cite{Haller2009}.

Our numerical results of \fref{maF2_single}  indicate that the physical mass monotonically decreases from a value of $-m$ in the weakly interacting to $-2m$ in the strongly attractive regime. The inertial mass, even though it takes the same limiting values, appears to have a faint maximum, reaching values of $-0.990m$ around $\gamma \approx -0.7$. An interesting regime  is the crossover around $\gamma \approx -1$ where the mass ratio takes a minimum value of ${m_I}/{m_P} \approx 0.78$ around $\gamma\approx -1.3$ while it stays below unity for the whole interaction range.

The negative physical mass and missing particle number $N_d$ confirm that the yrast states correspond to hole-like or dark-soliton-like excitations even though the dimer-dimer coupling is negative (see \sref{sTG}). This may seem surprising since attractive interactions usually give rise to bright solitons. For the attractive Yang-Gaudin gas (and by equivalence the super Tonks-Girardeau gas), however, the compressibility is positive in the whole interaction regime  [see equation (\ref{speedofsound})], which implies that it is stable against collapse into bright solitons. The positive compressibility is a consequence of Fermi pressure for the Yang Gaudin gas, and a corresponding kinetic energy pressure for the super Tonks-Girardeau gas. A manifestation of the effectively attractive interaction between dimers is the fact that the physical and inertial mass parameters are reduced compared to the value of the bare dimer hole mass $-2m$. 
\section{\label{disc_conc}Discussion and conclusions}
In this work we have identified the yrast excitations of the spin-$\frac{1}{2}$ Fermi gas with attractive $\delta$ interactions in the crossover from the dimerized regime of strong attraction to the non-interacting limit. We have further examined properties of the yrast dispersion relation in the thermodynamic limit obtained from the exact Bethe ansatz solutions of the Yang-Gaudin model. These predictions can also be directly applied to the super Tonks-Girardeau gas, on account of its mathematical equivalence to the fully dimerized solutions of the Yang-Gaudin model. The yrast dispersion relations very much resemble those of dark solitons in the nonlinear Schr\"odinger equation and the yrast dispersion of the Lieb-Liniger model of the one-dimensional Bose gas. Indeed, by the nature of the yrast states being the lowest energy state at given momentum, we can expect these excitations to be robust and long lived. Yet there are important quantitative differences between the results of the current work and the repulsive Bose gas. For example, the missing particle number for yrast excitations in the Bose gas can be macroscopically large as $N_d\sim -2 \gamma^{-1/2}$ diverges for small values of the interaction parameter $\gamma \gtrsim 0$ of the Lieb-Liniger model \cite{Astrakharchik2012}, whereas in the Yang-Gaudin gas $N_d > -2$ indicates that hole excitations are microscopic and smaller than a single bare dimer. Also, in contrast to the Bose gas where the effective mass ratio $m_I/m_P$ is always larger than unity \cite{Astrakharchik2012}, the ratio falls below one for the Yang-Gaudin gas, which means that yrast quasiparticles would oscillate with a frequency that is slightly faster than the trap frequency. These interesting effects are challenging to observe in experiments, since a one-dimensional Fermi gas would have to be realized and manipulated at the level of single atoms. Important steps have already been taken by realizing one-dimensional Fermi gases \cite{Pagano2014,Zurn2013,Serwane2011} and single-
particle-level detection with high-numerical aperture imaging \cite{Cheuk2015,Haller2015}.

A possible route towards measuring the oscillation frequency of the yrast quasiparticles in the Yang-Gaudin regime might be provided by the connection of the Yang-Gaudin with the Lieb-Liniger regimes in the one-dimensional gas across the confinement-induced resonance \cite{Fuchs2004,Tokatly2004,Astrakharchik2004b} following loosely the procedure applied in  \cite{Yefsah}: Yrast excitations could be excited by phase imprinting \cite{Denschlag2000} or combined phase and density imprinting \cite{Carr2001} and detected through a visible density depletion in the weakly-interacting Lieb-Liniger regime for tightly-bound dimers where dark solitons become macroscopic. A rapid ramp of interaction strength through the confinement-induced resonance would then adiabatically transform the dark solitons into the yrast states of the Yang-Gaudin model, thus allowing them to be probed. A similar procedure is feasible with a one-dimensional Bose gas with the aim of probing the yrast excitations of the super Tonks-Girardeau gas. A ramp through the confinement-induced resonance into the super Tonks-Girardeau regime has already been demonstrated \cite{Haller2009}.
\ack
The authors thank Martin Zwierlein, Victor Galitski, Xia-Ji Liu, and Hui Hu for helpful discussions and in particular Dimitri Gangardt for pointing out \cite{Schecter2012} and the thesis \cite{Campbell2013}.
\appendix
\section{\label{TechDetails}Finite system equations: technical details}
In this appendix we provide all the details on how the Bethe ansatz equations were solved numerically. We used the MATLAB environment, and in particular the \texttt{fsolve} function, included in the optimization toolbox.

In order to solve any of the logarithmic equations (where all variables are necessarily real and distinct), we always begin from the strong-coupling limit, taken as $|\gamma|=100$. In this regime, one can use $2\pi/L$-multiples of the quantum numbers as an initial guess, and the solver easily picks up the correct solution. Now, usually this guess is sufficient at any $\gamma$ (with the logarithmic equations), but we find it is better to follow the solutions from $|\gamma|=100$ down to whatever value one needs. This is done with an adapting step of $\gamma/10$, since in the strong-coupling regime the solutions change very little and large steps can be taken, while in the weak-coupling regime the converse is true. At each consecutive $\gamma$-step, the guess is taken as the solution at the previous step. This following-in-$\gamma$ procedure greatly improves efficiency and accuracy for large systems ($N \sim 100-1000$) but is optional for small systems ($N\sim 10$).

Now, solving the exact exponential Bethe ansatz equations with complex rapidities is a somewhat more involved task. Here there are two possibilities: either the structure of the solution changes with $\gamma$ (for example, distinct real rapidities may merge into a complex-conjugate pair), or it does not. In the latter case, the situation is simpler, so we start there. The simplest of all cases is the ground state, due to the symmetry of the solution. As such, we can track the solution somewhat further than for excited states.

For the ground state we begin from the known approximate solutions at $\gamma\approx0$ and track in $\gamma$ using a fixed step of $0.01$. Once again, we use a linear interpolation based on the previous two points as a guess for the next. Whenever the solver fails to converge to a valid solution, we call the solver a maximum of five times more, at each attempt adding a vector of (real) random numbers to the initially-used guess of order $10^{-4}$. If all five attempts are exhausted unsuccessfully, the program aborts.

For the single fermion branch, the structure of the solution changes as a function of $\gamma$. As always, we begin from $\gamma=-0.01$ where solutions are approximately known, use a fixed step of $0.01$ (unless we are in the vicinity of a merging -- see below) and linear interpolation for predictive guessing. At each step, the solver is called once and the solutions are inspected. The Bethe ansatz equations must have distinct roots, so that no two $k_j$'s and no two $\alpha_m$'s are equal. At points where two real rapidities merge into a complex-conjugate pair, the solver often fails to automatically split up these roots along the imaginary axis, so this must be tested for and corrected.

Thus, having made an initial attempt to solve at a given $\gamma$ value, we test whether any of the $k_j$'s have been returned equal (in this case they will also necessarily be real). If so, we modify the guess by splitting up these rapidities by $\pm0.1i n_0$. Having modified the guess to help the solver pick up the correct solution, we call it again and reduce step size to $0.001$ in a $\gamma$-interval of $0.01$ immediately following the merging point. The step is reduced because usually variables change very rapidly in the vicinity of a merging and in order for our linear guess to be effective, the $\gamma$-step must be smaller. Once we perform $10$ of these reduced steps, the step size is returned to normal.

Regardless of whether a merging takes place at any given $\gamma$ value, if the solver fails to converge, we make a maximum of ten further attempts to solve by adding small (real) random numbers of order $10^{-4} n_0$ to the initial guess (which may have already been ``manually'' modified due to a merging). If all ten attempts fail, the program is aborted.
\section{\label{TDLhowto}Thermodynamic limit equations: outline of derivation}
Since the ground state thermodynamic limit equations of the attractive Yang-Gaudin model are available in the literature (e.g.~\cite{Guan2011}), we will not explain how these are to be obtained from the finite-system discrete equations. However, we will indicate how one can obtain the excited-state equations starting from the ground state.

In the case of single fermion holes (which in the thermodynamic limit are dimer holes), we begin from the most convenient ground state, that of a system with $M+1$ dimers. This chosen initial ground state is then perturbed by explicitly removing a dimer to create a hole excitation. This perturbation to the system shifts all other rapidities, and this shifted ground state will be denoted by an asterisk. The explicit removal of a dimer, together with the shift of all other rapidities, will yield the excited state of interest. This excited state must be compared to the initial ground state from which we constructed it, which is not be the ground state of interest (which has $M$ dimers). We must then correct for the ``wrong'' ground state explicitly, which is a much simpler task then beginning from the correct but inconvenient ground state.

In our notation, the equations for the single fermion hole branch are derived by following the recipe:
\begin{equation}
\left\{\left[E^{\ast}(M+1)-1\alpha\right]-E(M+1)\right\}+ \left\{E(M+1)-E(M)\right\}.
\end{equation}
This means we must take the finite-system Bethe ansatz equations \footnote{Note that one must use the equations in logarithmic form and if applicable, under the string hypothesis.} of $M+1$ dimers, remove one charge rapidity (denoted here by $1\alpha$) with the other rapidities shifted from their true ground state values (denoted by the asterisk) and compare the resulting system to the true ground state of $M+1$ dimers. Finally, we add a ground state correction that ensures that both the excited and the ground states have $N=2M$ fermions (in practice, the ground state correction simply contributes a $2\mu$ term to the excitation energy since the energy cost of adding/removing two fermions to the ground state is precisely $2\mu$).

Now let us outline in more detail how the equations are derived: the shift in $\alpha_m$ due to the perturbation is written as $\delta \alpha_m\ L$. We write down the finite-system Bethe ansatz equations for the chosen initial ground state, shifting all the rapidities. Then we explicitly subtract any interaction terms that are associated with the rapidities that must be removed, which are themselves not shifted. Next, we expand the interaction terms [in the case of the $\delta$-function potential, the two-body phase-shifts are given by the $\theta$ function of (\ref{thetafnct})] as a first order Taylor expansion about the unshifted rapidities, with the first order terms proportional to the shifts.

We now subtract the Bethe ansatz equations of the unperturbed, conveniently-chosen ground state. At this point we may pass to the continuous limit by using the ground state quasi-momenta density function that appear in the ground state thermodynamic limit equations with the following rule:
\begin{eqnarray}
\sum\limits_{m}\rightarrow L\int\limits_{-b}^b r(\alpha)d\alpha. 
\end{eqnarray}
Some of the terms appearing under the integrals in the equations can be simplified by substituting for them from the ground state thermodynamic limit integral equations. Finally, we define a new function that is the product of $L$, the ground state distribution, and the shift: in particular $J(\alpha)=L\  \delta\alpha\ r(\alpha)$.

It remains to compute the energy and momentum. This is done according to our recipe, by writing down the finite-system total $P$ and $E$ of the excited state (shifted and with an explicit rapidity removed), subtracting those of the convenient ground state, and adding the ground state correction. The resulting quantities can be expressed through the ``shift function'' of the integrals equations (i.e.~$J(\alpha)$), the explicit rapidity removed and the chemical potential.
\section{\label{Ns_aF2s}Evaluating $N_d$}
Here we give details on how we evaluate (\ref{Schectdef}) for the yrast excitation in the attractive Yang-Gaudin model directly. We first write this equation in the equivalent form
\begin{equation}
\label{Schectdef2}
N_d = \frac{-\left.\frac{dE}{d\mu}\right|_{v_s} + v_s \left.\frac{dP}{d\mu}\right|_{v_s} - \frac{v_s}{n_0}\frac{dn_0}{d\mu}P}{1-\frac{mv_s^2}{n_0}\frac{dn_0}{d\mu}},
\end{equation}
where momentum was replaced by the velocity $v_s$ as an independent variable. Define
\begin{eqnarray}
\label{ffncts1}
f_1(\gamma) &=& \left[\int\limits_{-1}^1 g(x) dx\right]^2,\\
f_2(\gamma,q) &=& 2\int\limits_{-1}^1 xh(x) dx,\\
f_3(\gamma,q) &=& \int\limits_{-1}^1 h(x) dx,
\label{ffncts3}
\end{eqnarray}
and write the excitation energy and momentum through the scaled variables as
\begin{eqnarray}
E &=& \frac{\hbar^2}{2m}\frac{n_0^2}{2 f_1(\gamma)} \left[-q^2+ f_2(\gamma,q) \right] + 2\mu,\\
P &=& \hbar\frac{n_0}{\sqrt{f_1(\gamma)}} \left[-q+ f_3(\gamma,q) \right].
\label{eqnPc}
\end{eqnarray}
The velocity is defined as
\begin{equation}
v_s\equiv\frac{dE}{dP}=\frac{dE}{dq}\frac{dq}{dP}=\frac{\hbar}{m}\frac{n_0}{2\sqrt{f_1}}\left\{\frac{-2q+\frac{df_2}{dq}}{-1+\frac{df_3}{dq}}\right\}.
\label{speed}
\end{equation}
To begin with, we need to take the derivative of $E$ with respect to $\mu$:
\begin{eqnarray}
\label{E_deriv}
\frac{dE}{dn_0} &=& \frac{\hbar^2}{2m}\frac{2n_0f_1-n_0^2\frac{df_1}{dn_0}}{2f_1^2} \left[-q^2+ f_2\right] \nonumber\\
&+& \frac{\hbar^2}{2m}\frac{n_0^2}{2f_1}\left(-2q\frac{dq}{dn_0}+\frac{df_2}{dn_0}\right)+2\frac{d\mu}{dn_0},\\
\frac{dE}{d\mu} &=& \frac{dE}{dn_0} \frac{dn_0}{d\mu}.
\end{eqnarray}
Differentiating the chemical potential, we get
\begin{eqnarray}
\frac{d\mu}{dn_0} &=&\frac{\hbar^2}{2m} \left[2n_0\alpha(\gamma) - c\beta(\gamma)\right],\\
\alpha(\gamma) &=& 3e(\gamma)-\gamma\frac{de(\gamma)}{d\gamma},\\
\beta(\gamma) &=& 2\frac{de(\gamma)}{d\gamma}-\gamma\frac{d^2e(\gamma)}{d\gamma^2}.
\end{eqnarray}
The derivatives we need are
\begin{eqnarray}
\frac{df_1}{dn_0} &=& \frac{df_1}{d\gamma}\frac{d\gamma}{dn_0},\\
\frac{df_2}{dn_0} &=& \frac{df_2}{d\gamma}\frac{d\gamma}{dn_0} + \frac{df_2}{dq}\frac{dq}{dn_0}.
\end{eqnarray}
Next, we need $dq/dn_0$ keeping $v_s$ constant and we have no closed analytical expression for $q$. Still, this derivative can be done. Defining $\tilde{v}_s=v_s/n_0$ and going back to (\ref{speed}):
\begin{equation}
\label{eqn30}
\tilde{v}_s = \frac{v_s}{n_0} = \frac{\hbar}{m}\frac{1}{\sqrt{2f_1}}\left\{\frac{-2q+\frac{df_2}{dq}}{-1+\frac{df_3}{dq}}\right\}.
\end{equation}
The right hand side only explicitly depends on $q$ and $\gamma$. Therefore, there exists some relation linking $\tilde{v}_s, q, \gamma$ which can in principle be solved to give $q(\tilde{v}_s,\gamma)$. In this case, we have
\begin{equation}
\frac{dq}{dn_0} = \frac{dq}{d\gamma}\frac{d\gamma}{dn_0}+\frac{dq}{d\tilde{v}_s}\frac{d\tilde{v}_s}{dn_0}.
\end{equation}
The difficult derivative appearing here is $\frac{dq}{d\gamma}$ keeping $v_s$ constant. Since $\tilde{v}_s$ is only a function of $\gamma$ and $q$, we can write
\begin{equation}
d\tilde{v}_s = \frac{\partial \tilde{v}_s}{\partial \gamma}d\gamma + \frac{\partial \tilde{v}_s}{\partial q}dq.
\end{equation}
We want to keep $\tilde{v}_s$ constant, so $d\tilde{v}_s$ must be zero, from which we get
\begin{equation}
\frac{dq}{d\gamma} = -\frac{\partial \tilde{v}_s}{\partial \gamma}\frac{\partial q}{\partial \tilde{v}_s}.
\end{equation}
All together, we have
\begin{equation}
\frac{dq}{dn_0} = \frac{1}{n_0^2}\left(c\frac{\partial \tilde{v}_s}{\partial \gamma}-v_s\right)\frac{\partial q}{\partial \tilde{v}_s}.
\end{equation}
$\frac{\partial \tilde{v}_s}{\partial q}$ and $\frac{\partial \tilde{v}_s}{\partial \gamma}$ are to be found numerically, but since $\tilde{v}_s$ is defined by the right hand side of (\ref{eqn30}), this can be done. We now have all the necessary equations to compute $\left.\frac{dE}{d\mu}\right|_{v_s}$.

Next, we need to compute $\left.\frac{dP}{d\mu}\right|_{v_s}$. Taking the derivative of (\ref{eqnPc}):
\begin{eqnarray}
\label{P_deriv}
\frac{dP}{dn_0} &=& \frac{\hbar}{\sqrt{f_1}}\left(f_3-q\right) - \frac{\hbar n_0}{2}f_1^{-3/2} \frac{df_1}{dn_0} \left(f_3-q\right)\nonumber\\
&+& \frac{\hbar n_0}{\sqrt{f_1}} \left(\frac{df_3}{dn_0} - \frac{dq}{dn_0} \right),\\
\frac{dP}{d\mu} &=& \frac{dP}{dn_0} \frac{dn_0}{d\mu}.
\end{eqnarray}
The only new derivative appearing here is
\begin{equation}
\frac{df_3(\gamma,q)}{dn_0} = \frac{df_3}{d\gamma}\frac{d\gamma}{dn_0} + \frac{df_3}{dq}\frac{dq}{dn_0}.
\end{equation}
We now have everything we need to calculate $N_d$ from (\ref{Schectdef2}) directly.
\section*{References}
\bibliography{ourbib2,Sophie1}

\providecommand{\newblock}{}
\begin{thebibliography}{10}
\expandafter\ifx\csname url\endcsname\relax
  \def\url#1{{\tt #1}}\fi
\expandafter\ifx\csname urlprefix\endcsname\relax\def\urlprefix{URL }\fi
\providecommand{\eprint}[2][]{\url{#2}}

\bibitem{Dauxois2006}
Dauxois T and Peyrard M 2006 {\em Physics of Solitons\/} (Cambridge University
  Press)

\bibitem{Kevrekidis2008}
Kevrekidis P~G, Frantzeskakis D~J and Carretero-Gonz{\'{a}}lez R (eds) 2008
  {\em Emergent Nonlinear Phenomena in Bose-Einstein Condensates\/} ({\em
  Atomic, Optical, and Plasma Physics\/} vol~45) (Berlin, Heidelberg: Springer
  Berlin Heidelberg)

\bibitem{Tsuzuki1971}
Tsuzuki T 1971 {\em J. Low Temp. Phys.\/} {\bf 4} 441

\bibitem{Burger1999}
Burger S, Bongs K, Dettmer S, Ertmer W and Sengstock K 1999 {\em Phys. Rev.
  Lett.\/} {\bf 83} 5198--5201

\bibitem{Denschlag2000}
Denschlag J, Simsarian J~E, Feder D~L, Clark C~W, Collins L~A, Cubizolles J,
  Deng L, Hagley E~W, Helmerson K, Reinhardt W~P, Rolston S~L, Schneider B~I
  and Phillips W~D 2000 {\em Science\/} {\bf 287} 97--101

\bibitem{Becker2008}
Becker C, Stellmer S, Soltan-Panahi P, D{\"{o}}rscher S, Baumert M, Richter
  E~M, Kronj{\"{a}}ger J, Bongs K and Sengstock K 2008 {\em Nat. Phys.\/} {\bf
  4} 496--501

\bibitem{Kinast2005}
Kinast J, Turlapov A, Thomas J~E, Chen Q, Stajic J and Levin K 2005 {\em
  Science\/} {\bf 307} 1296--1299

\bibitem{Zwierlein2005}
Zwierlein M~W, Abo-Shaeer J~R, Schirotzek A, Schunck C~H and Ketterle W 2005
  {\em Nature\/} {\bf 435} 1047--1051

\bibitem{Liao2010}
Liao Y~a, Rittner A~S~C, Paprotta T, Li W, Partridge G~B, Hulet R~G, Baur S~K
  and Mueller E~J 2010 {\em Nature\/} {\bf 467} 567--9

\bibitem{Navon2010}
Navon N, Nascimb{\`{e}}ne S, Chevy F and Salomon C 2010 {\em Science\/} {\bf
  328} 729--32

\bibitem{Joseph2011}
Joseph J~A, Thomas J~E, Kulkarni M and Abanov A~G 2011 {\em Phys. Rev. Lett.\/}
  {\bf 106} 150401

\bibitem{Giorgini2008}
Giorgini S and Stringari S 2008 {\em Rev. Mod. Phys.\/} {\bf 80} 1215--1274

\bibitem{Zwerger2012}
Zwerger W (ed) 2012 {\em The BCS –- BEC Crossover and the Unitary Fermi
  Gas\/} vol 836 (Springer Berlin / Heidelberg,)

\bibitem{Dziarmaga2004b}
Dziarmaga J and Sacha K 2004 Soliton in {BCS} superfluid {F}ermi gas
  (\textit{Preprint} \eprint{arXiv:0407585})

\bibitem{Antezza2007}
Antezza M, Dalfovo F, Pitaevskii L and Stringari S 2007 {\em Phys. Rev. A\/}
  {\bf 76} 043610

\bibitem{Liao11pr:FermiSolitons}
Liao R and Brand J 2011 {\em Phys. Rev. A\/} {\bf 83} 041604(R)

\bibitem{Scott2011}
Scott R, Dalfovo F, Pitaevskii L and Stringari S 2011 {\em Phys. Rev. Lett.\/}
  {\bf 106} 185301

\bibitem{Spuntarelli2011}
Spuntarelli A, Carr L~D, Pieri P and Strinati G~C 2011 {\em New J. Phys.\/}
  {\bf 13} 035010

\bibitem{Scott2012}
Scott R~G, Dalfovo F, Pitaevskii L~P, Stringari S, Fialko O, Liao R and Brand J
  2012 {\em New J. Phys.\/} {\bf 14} 023044

\bibitem{Ku}
Ku M~J~H, Mukherjee B, Yefsah T and Zwierlein M~W 2016 {\em Phys. Rev. Lett.\/}
  {\bf 116} 045304

\bibitem{Xu2014}
Xu Y, Mao L, Wu B and Zhang C 2014 {\em Phys. Rev. Lett.\/} {\bf 113} 130404

\bibitem{Liu2015}
Liu X~J 2015 {\em Phys. Rev. A\/} {\bf 91} 023610

\bibitem{Zou2015}
Zou P, Brand J, Liu X~j and Hu H 2015 {\em arXiv Prepr.\/} (\textit{Preprint}
  \eprint{1509.01803}) \urlprefix\url{http://arxiv.org/abs/1509.01803}

\bibitem{Konotop2004}
Konotop V~V and Pitaevskii L 2004 {\em Phys. Rev. Lett.\/} {\bf 93} 240403

\bibitem{Busch2000}
Busch T and Anglin J~R 2000 {\em Phys. Rev. Lett.\/} {\bf 84} 2298--2301

\bibitem{Weller2008}
Weller A, Ronzheimer J~P, Gross C, Esteve J, Oberthaler M~K, Frantzeskakis D~J,
  Theocharis G and Kevrekidis P~G 2008 {\em Phys. Rev. Lett.\/} {\bf 101}
  130401

\bibitem{Cetoli2013}
Cetoli A, Brand J, Scott R~G, Dalfovo F and Pitaevskii L~P 2013 {\em Phys. Rev.
  A\/} {\bf 88} 043639

\bibitem{Mateo2014}
Mu{\~{n}oz Mateo} A and Brand J 2014 {\em Phys. Rev. Lett.\/} {\bf 113} 255302

\bibitem{brand01a}
Brand J and Reinhardt W~P 2001 {\em J. Phys. B At. Mol. Opt. Phys.\/} {\bf 34}
  L113--L119

\bibitem{Brand2002}
Brand J and Reinhardt W~P 2002 {\em Phys. Rev. A\/} {\bf 65} 043612

\bibitem{Komineas2003}
Komineas S and Papanicolaou N 2003 {\em Phys. Rev. A\/} {\bf 68} 043617

\bibitem{Yefsah}
Yefsah T, Sommer A~T, Ku M~J~H, Cheuk L~W, Ji W, Bakr W~S and Zwierlein M~W
  2013 {\em Nature\/} {\bf 499} 426--30

\bibitem{Ku2014}
Ku M~J~H, Ji W, Mukherjee B, Guardado-Sanchez E, Cheuk L~W, Yefsah T and
  Zwierlein M~W 2014 {\em Phys. Rev. Lett.\/} {\bf 113} 065301

\bibitem{Mateo2015a}
Mu{\~{n}oz Mateo} A and Brand J 2015 {\em New J. Phys.\/} {\bf 17} 125013

\bibitem{Muryshev1999}
Muryshev A~E, {van Linden van den Heuvell} H~B and Shlyapnikov G~V 1999 {\em
  Phys. Rev. A\/} {\bf 60} R2665--R2668

\bibitem{Efimkin2014}
Efimkin D~K and Galitski V 2015 {\em Phys. Rev. A\/} {\bf 91} 023616

\bibitem{Mermin1966}
Mermin N~D and Wagner H 1966 {\em Phys. Rev. Lett.\/} {\bf 17} 1133--1136

\bibitem{hohenberg67}
Hohenberg P~C 1967 {\em Phys. Rev.\/} {\bf 158} 383--386

\bibitem{Liu2007b}
Liu X~J, Hu H and Drummond P~D 2007 {\em Phys. Rev. A\/} {\bf 76} 043605

\bibitem{Yang}
Yang C~N 1967 {\em Phys. Rev. Lett.\/} {\bf 19}(23) 1312--1315

\bibitem{Gaudin}
Gaudin M 1967 {\em Physics Letters A\/} {\bf 24} 55--56

\bibitem{Fuchs2004}
Fuchs J~N, Recati A and Zwerger W 2004 {\em Phys. Rev. Lett.\/} {\bf 93} 090408

\bibitem{Tokatly2004}
Tokatly I~V 2004 {\em Phys. Rev. Lett.\/} {\bf 93} 090405

\bibitem{Astrakharchik2004b}
Astrakharchik G, Blume D, Giorgini S and Pitaevskii L 2004 {\em Phys. Rev.
  Lett.\/} {\bf 93} 050402

\bibitem{Mora2005}
Mora C, Komnik A, Egger R and Gogolin A~O 2005 {\em Phys. Rev. Lett.\/} {\bf
  95} 080403

\bibitem{Bergeman2003}
Bergeman T, Moore M~G and Olshanii M 2003 {\em Phys. Rev. Lett.\/} {\bf 91}
  163201

\bibitem{LL1}
Lieb E~H and Liniger W 1963 {\em Phys. Rev.\/} {\bf 130}(4) 1605--1616

\bibitem{LL2}
Lieb E~H 1963 {\em Phys. Rev.\/} {\bf 130}(4) 1616--1624

\bibitem{olshanii98}
Olshanii M 1998 {\em Phys. Rev. Lett.\/} {\bf 81} 938--941

\bibitem{girardeau60}
Girardeau M 1960 {\em J. Math. Phys.\/} {\bf 1} 516--523

\bibitem{Kanamoto2008}
Kanamoto R, Carr L and Ueda M 2008 {\em Phys. Rev. Lett.\/} {\bf 100} 060401

\bibitem{Kulish1976}
Kulish P~P, Manakov S~V and Faddeev L~D 1976 {\em Theor. Math. Phys.\/} {\bf
  28} 615--620

\bibitem{Ishikawa1980}
Ishikawa M and Takayama H 1980 {\em J. Phys. Soc. Japan\/} {\bf 49} 1242

\bibitem{Kanamoto2009}
Kanamoto R, Carr L and Ueda M 2009 {\em Phys. Rev. A\/} {\bf 79} 1--12

\bibitem{Kanamoto2010}
Kanamoto R, Carr L~D and Ueda M 2010 {\em Phys. Rev. A - At. Mol. Opt. Phys.\/}
  {\bf 81} 023625

\bibitem{Fialko2012b}
Fialko O, Delattre M~C, Brand J and Kolovsky A 2012 {\em Phys. Rev. Lett.\/}
  {\bf 108} 250402

\bibitem{Syrwid2015}
Syrwid A and Sacha K 2015 {\em Phys. Rev. A\/} {\bf 92} 032110

\bibitem{Kaminishi2011}
Kaminishi E, Kanamoto R, Sato J and Deguchi T 2011 {\em Phys. Rev. A\/} {\bf
  83} 031601(R)

\bibitem{Sato2012a}
Sato J, Kanamoto R, Kaminishi E and Deguchi T 2012 {\em Phys. Rev. Lett.\/}
  {\bf 108} 110401

\bibitem{Sato2016}
Jun~Sato Rina~Kanamoto E~K~T~D 2016  (\textit{Preprint}
  \eprint{arXiv:1602.08329 [cond-mat.quant-gas]})

\bibitem{Astrakharchik2012}
Astrakharchik G~E and Pitaevskii L~P 2012 {\em EPL (Europhysics Letters)\/}
  {\bf 102} 30004

\bibitem{Astrakharchik2004}
Astrakharchik G~E, Blume D, Giorgini S and Granger B~E 2004 {\em Phys. Rev.
  Lett.\/} {\bf 92}(3) 030402

\bibitem{Astrakharchik2005}
Astrakharchik G~E, Boronat J, Casulleras J and Giorgini S 2005 {\em Phys. Rev.
  Lett.\/} {\bf 95}(19) 190407
  \urlprefix\url{http://link.aps.org/doi/10.1103/PhysRevLett.95.190407}

\bibitem{Haller2009}
Haller E, Gustavsson M, Mark M~J, Danzl J~G, Hart R, Pupillo G and N{\"a}gerl
  H~C 2009 {\em Science\/} {\bf 325} 1224--1227

\bibitem{Batchelor2005}
Batchelor M~T, Bortz M, Guan X~W and Oelkers N 2005 {\em Journal of Statistical
  Mechanics: Theory and Experiment\/} {\bf 2005} L10001
  \urlprefix\url{http://stacks.iop.org/1742-5468/2005/i=10/a=L10001}

\bibitem{Batchelor2010}
Chen S, Guan X~W, Yin X, Guan L and Batchelor M~T 2010 {\em Phys. Rev. A\/}
  {\bf 81}(3) 031608

\bibitem{Batchelor2005II}
Batchelor M~T, Bortz M, Guan X~W and Oelkers N 2006 {\em Journal of Physics:
  Conference Series\/} {\bf 42} 5

\bibitem{Sutherland2004}
Sutherland B 2004 {\em Beautiful Models: 70 Years of Exactly Solved Quantum
  Many-body Problems\/} (Singapore: World Scientific)

\bibitem{Sykes2007}
Sykes A~G, Drummond P~D and Davis M~J 2007 {\em Phys. Rev. A\/} {\bf 76}(6)
  063620

\bibitem{GaudinThesis}
Gaudin M 1967 {\em Etude d'un modele a une dimension pour un systeme de
  fermions en interaction\/} Ph.D. thesis Centre d'Etudes Nucleaires de Saclay

\bibitem{Guan2013}
Guan X~W~W, Batchelor M~T and Lee C 2013 {\em Rev. Mod. Phys.\/} {\bf 85}
  1633--1691

\bibitem{Guan2011}
Guan X~W and Ma Z~Q 2012 {\em Phys. Rev. A\/} {\bf 85}(3) 033632

\bibitem{Krivnov}
Krivnov V and Ovchinnikov A 1975 {\em Soviet Journal of Experimental and
  Theoretical Physics\/} {\bf 40} 781

\bibitem{Schlottmann}
Schlottmann P 1994 {\em Journal of Physics: Condensed Matter\/} {\bf 6} 1359

\bibitem{Ma2012}
Zhou L, Xu C~Y and li~Ma Y 2012 {\em Journal of Statistical Mechanics: Theory
  and Experiment\/} {\bf 2012} L03002

\bibitem{Atkinson2008}
Atkinson K~E and Shampine L~F 2008 {\em ACM Trans. Math. Softw.\/} {\bf 34} 21

\bibitem{pitaevskii03:book}
Pitaevskii L and Stringari S 2003 {\em Bose-Einstein Condensation\/} (Oxford:
  Clarendon)

\bibitem{Schecter2012}
Schecter M, Gangardt D~M and Kamenev A 2012 {\em Ann. Phys. (N. Y).\/} {\bf
  327} 639--670

\bibitem{Sophie:unpup}
Shamailov S~S and Brand J 2016 unpublished

\bibitem{petrov00}
Petrov D~S, Shlyapnikov G~V and Walraven J~T~M 2000 {\em Phys. Rev. Lett.\/}
  {\bf 85} 3745--3749

\bibitem{Campbell2013}
Campbell A~S 2013 {\em Mobile impurities in one-dimensional quantum liquids\/}
  Phd thesis University of Birmingham

\bibitem{Pagano2014}
Pagano G, Mancini M, Cappellini G, Lombardi P, Sch{\"{a}}fer F, Hu H, Liu X~j,
  Catani J, Sias C, Inguscio M and Fallani L 2014 {\em Nat. Phys.\/} {\bf 10}
  198--201

\bibitem{Zurn2013}
Z{\"{u}}rn G, Wenz A~N, Murmann S, Bergschneider A, Lompe T and Jochim S 2013
  {\em Phys. Rev. Lett.\/} {\bf 111} 175302

\bibitem{Serwane2011}
Serwane F, Z{\"{u}}rn G, Lompe T, Ottenstein T~B, Wenz a~N and Jochim S 2011
  {\em Science\/} {\bf 332} 336--338

\bibitem{Cheuk2015}
Cheuk L~W, Nichols M~A, Okan M, Gersdorf T, Ramasesh V~V, Bakr W~S, Lompe T and
  Zwierlein M~W 2015 {\em Phys. Rev. Lett.\/} {\bf 114} 193001

\bibitem{Haller2015}
Haller E, Hudson J, Kelly A, Cotta D~A, Peaudecerf B, Bruce G~D and Kuhr S 2015
  {\em Nat. Phys.\/} {\bf 11} 738--742

\bibitem{Carr2001}
Carr L, Brand J, Burger S and Sanpera a 2001 {\em Phys. Rev. A\/} {\bf 63}
  051601(R)

\end{thebibliography}
\end{document}